\documentclass[11pt]{article} 

\usepackage[margin=1in]{geometry} 
\usepackage{amsthm,amssymb, amsfonts, amsmath}
\usepackage{titling, fancyhdr}
\usepackage{enumerate}
\usepackage{graphicx}
\usepackage{color}
\usepackage{float}
\floatstyle{plaintop}
\restylefloat{table}
\usepackage{hyperref}
\usepackage{multicol}
\usepackage{mathtools}
\usepackage[dvipsnames]{xcolor}
\usepackage[justification=centering]{caption}
\usepackage{subfig}
\usepackage{booktabs}

\allowdisplaybreaks

\edef\restoreparindent{\parindent=\the\parindent\relax}
\usepackage{parskip}
\restoreparindent
\usepackage{indentfirst}

\usepackage{authblk}
\title{Nonproliferation and fusion power plants}
\author[1]{Michael Y. Hua}
\author[2]{Sachin S. Desai}
\author[3]{Amy C. Roma}
\author[4]{Angela Di Fulvio}
\author[5]{Craig J. Mundie}
\author[6]{Sara A. Pozzi}
\affil[1]{Department of Radiation Safety and Nuclear Science, Helion Energy, Everett, WA}
\affil[2]{Office of the General Counsel, Helion Energy, Everett, WA}
\affil[3]{Global Energy Practice, Hogan Lovells, Washington, D.C.}
\affil[4]{Department of Nuclear, Plasma \& Radiological Engineering, University of Illinois Urbana-Champaign, Champaign, IL}
\affil[5]{Mundie \& Associates, Seattle, WA}
\affil[6]{Department of Nuclear Engineering \& Radiological Sciences, University of Michigan, Ann Arbor, MI}
\date{\today}

\begin{document}
	\maketitle 

\section*{Abstract}
As fusion energy progresses towards commercial deployment, the question has arisen as to the role of the Treaty on the Non Proliferation of Nuclear Weapons (NPT) with respect to fusion, including whether the nuclear nonproliferation regime includes -- or should be extended to include -- future fusion plants. 

The paper first addresses the current nonproliferation regime and its application to fusion.  This regime, solidified in and based on the NPT, is designed to ensure that certain types of radioactive material and technology are used only for peaceful purposes and not nuclear weapons purposes.  Specifically, the NPT expressly controls the transfer of source material (i.e., unenriched uranium and thorium); special fissionable material (i.e., enriched uranium (U-235), uranium-233, and plutonium-239); and equipment that is ``especially designed or prepared for processing, use, or production of special fissionable material."  

As spelled out in the plain language of the NPT and in the NPT's implementing documents, the scope of the NPT-based nonproliferation regime -- with its focus on safeguards -- is limited to fission-related technology, including fission reactors and the fission nuclear fuel cycle technologies (e.g., enrichment and conversion facilities). As fusion does not use source or special fissionable material or the fission nuclear fuel cycle, commercial fusion applications \textit{prima facie} fall outside the current NPT nonproliferation regime and implementing documents. 

This paper then addresses options to consider for fusion, including whether fusion power plants \textit{should} be included within the NPT and associated safeguards framework, or whether other frameworks -- particularly the existing \textit{global}, dual-use export control framework -- are more appropriate to control the technology. Including fusion power plants within the NPT would requirement amendment of the treaty, which is largely a geopolitical issue and attaining the required global consensus is likely not feasible.  In any event, a technical analysis of fusion is considered.

Based on a technical analysis explained herein, and as applied to the existing legal framework, this paper concludes that commercial fusion facilities should continue to fall outside the NPT.  Including fusion and applying safeguards is not warranted considering that the nonproliferation risks associated with traditional nuclear fission just do not exist for fusion and would be beyond the ``extent necessary to ensure its use only for peaceful purposes." Rather, this paper concludes that fusion facilities as they are ``especially designed or prepared" have very limited significance to the development of special fissionable material for weapons purposes or nuclear weapons themselves. Any potential, malicious misuse of the fusion technology from a nonproliferation perspective would require significant material changes to the underlying technology and apparatus. Additionally, adding fusion to the NPT would not provide a gain commensurate with the incredible work required to amend the NPT and the associated implementing documents.  

In the alternative, this paper concludes that the current dual-use export control regime is an appropriate path to look to for fusion, in large part because any potentially significant malicious misuse of the fusion technology from a nonproliferation perspective would require significant material changes to the underlying technology by rogue actors. Applying and modifying a dual-use export controls approach as necessary alongside the NPT –- including potentially developing a ``controls by design'' usage-based controls regime for fusion -- can more effectively support the safe deployment of this essential technology rather than applying the ill-fit, fission-specific nonproliferation regime.

\pagebreak

\section{Introduction}
Research and development for fusion power plants has accelerated in recent years, and there has been a significant increase in private funding. This progress has been made possible by significant advancements in enabling technologies like modeling, power electronics, and magnet technology.  Multiple companies are seeking to demonstrate net energy (and in at least one case, even net electricity) from fusion within the next few years, with first power plants to follow~\cite{FIA_survey_2021,FIA_survey_2022,PR_Helion_1,PR_Helion_2,PR_Helion_3,PR_CFS_1,PR_CFS_2,PR_GF_1,PR_TAE_1,PR_TAE_2,PR_TE_1,PR_Zap_1,PR_firstlight_1}.

As fusion energy progresses towards near-term commercial deployment, the question arises as to the role of the Treaty on the Non Proliferation of Nuclear Weapons (NPT) with respect to fusion, and whether (i) the nuclear nonproliferation regime includes, or (ii) should be extended to include, future fusion plants.  Note that there exist some fusion-fission hybrid technologies, but these are out of the scope of this paper and not considered fusion power plants as they are still dependent on fission. In any event, their fission applications would be subject to the existing NPT framework. 

This paper analyses those two questions under the outline as follows. 
\begin{itemize}
	\item Section~\ref{sec:BG} addresses the current application of the NPT and global nonproliferation regime to fusion. To that end, it explains the historical and legal landscape for the global nonproliferation regime and analyzes the NPT's scope, IAEA safeguards agreements and supporting instruments, and the broader export control framework that sits alongside the NPT. 
	\begin{itemize}
		\item Section~\ref{sec:history} provides historical background information from World War II leading to the NPT
		\item Section~\ref{sec:scope} provides an overview of the NPT
		\item Section~\ref{sec:safeguards} explains role of the IAEA, safeguards agreements and supporting instruments
		\item Section~\ref{sec:export_controls} provides an overview of the export control framework
		\item Section~\ref{sec:fusion_spec} applies fusion to the NPT, IAEA and safeguards agreements, and the export control framework
	\end{itemize}
	\item Section~\ref{sec:meat} evaluates, given the understanding that the global nonproliferation regime as is does not apply to fusion, whether it should be amended to apply to fusion or if another approach (such as turning to export controls) would be more appropriate.  To that end, this section evaluates the different theoretical risks that have been identified related to fusion, their connection to nuclear proliferation, and what are the best tools to address those risks. These theoretical risks include:
	\begin{itemize}
		\item Section~\ref{sec:SNM}: use of fusion devices as a neutron source to create fissile material,
		\item Section~\ref{sec:knowledge}: leveraging research into ignited, inertially confined plasmas towards nuclear weapons development, and
		\item Section~\ref{sec:nukes}: diversion of tritium from fusion facilities to boost nuclear weapons. 
	\end{itemize}
	The analysis in Section~\ref{sec:meat} demonstrates that these theoretical concerns do not raise a material practical risk of nuclear weapons proliferation, and thus do not warrant extending a fission-focused nonproliferation regime (including safeguards) to fusion. Instead, an approach focused on export controls appears more appropriate. Note that diversion of tritium from fusion facilities towards a radiological dispersion device (RDD or ``dirty bomb") is ancillary to this discussion and addressed in Appendix~\ref{sec:RDD}. 
	
	\item Section~\ref{sec:conclusion} concludes this paper.
\end{itemize}

\section{Background: historical and legal landscape}\label{sec:BG}
\subsection{Historical context}\label{sec:history}
The dawn of the nuclear age at the end of the World War II introduced to the world two things: nuclear weapons, and soon thereafter nuclear power for peaceful use. When the United Nations was formed, shortly after the US deployed two nuclear weapons in Japan, its first order of business was to address the issues of preventing the spread of nuclear weapons while ensuring the peaceful use of nuclear technology.  The first UN resolution was to establish a commission to handle atomic energy~\cite{UN_1}, and set forth the general global consensus that nations (i) share scientific findings related to atomic energy, (ii) eliminate weapon of mass destruction arsenals, (iii) use safeguards to protect compliant states from the hazards of noncompliance, and (iv) have ``control of atomic energy to the extent necessary to ensure its use only for peaceful purposes"~\cite{UN_1}. 

In the subsequent couple decades post-World War II, particularly the 1950s and 1960s, the world saw a number of developments on both the nuclear weapons and peaceful use front.  On the weapons side, a number of countries, in addition to the US, successfully tested nuclear weapons, including the then-USSR, the United Kingdom, France, and China~\cite{ICAN}.  The first thermonuclear weapon, a second-generation nuclear weapon that uses a combination fission and fusion system, was tested in the early 1950s, and soon thereafter widely under development by nuclear weapons states. Notably, however, these weapons intrinsically rely on a fission process~\cite{LANL_primer,glasstone,krane}.

On the peaceful-use side, the first electricity production from a fission reactor occurred in 1951.  In 1953 President Dwight Eisenhower delivered his ``Atoms for Peace" speech to the United States, under which the US agreed to share nuclear technology with the rest of the world, in exchange for the rest of the world foregoing the development of nuclear weapons.  In 1957, the first reactor to connect with the grid became operational.  Soon therefore, civilian nuclear technology spread around the world, and dozens of countries have hundreds of nuclear reactors for peaceful uses, including power generation, medical use, and research and development~\cite{history_reactor}.

Fusion research and development was around at this time as well. By the World War II timeframe, just before the middle of the 20th century, the theoretical framework for fusion had been established, and fundamental science was still being explored. Fusion applications were being explored broadly in the US, Russia, and UK, while Japan, France, and Sweden were looking at developing fusion power plants. Whereas fission is the splitting of large nuclei ($\approx$240 atomic mass units), fusion is the combining of small nuclei ($\approx$2 atomic mass units). The domains of these reactions are on opposite ends of the periodic table: the uranium of fission is the last stable element, whereas the hydrogen of fusion is the first. Fission assemblies support self-sustaining chain reactions to produce energy, whereas fusion power plants do not use chain reactions to generate energy.

Given these developments, the global community needed to develop a global framework to prevent the spread of nuclear weapons, while enabling peaceful uses -- which the world wanted.  This is the foundation of the NPT, the cornerstone of the global nonproliferation regime~\cite{NPT}.

In 1952, the General Assembly created what is now known as United Nations Disarmament Commission (UNDC) under the Security Council with a mandate to prepare proposals for a treaty for the regulation, limitation, and balanced reduction of, among other things, nuclear weapons~\cite{UNDC}.  While only meeting occasionally in its early years, an earlier version of the UNDC, the Eighteen-Nation Disarmament Committee (ENDC), was formed in 1962. The ENDC was a committee that would ultimately negotiate the NPT\footnote{The UNDC undertook a few evolutions, with earlier versions including the ENCD, until it ultimately settled in its current form, the UNDC, in 1968.  The UN General Assembly accepted the decision of the major powers to create the Eighteen Nation Committee on Disarmament (ENCD) through resolution 1722 (XVI) on December 21, 1961.}~\cite{ENDC_1}.   

The ENCD, was made up of 18 nations: Canada, France (in a non-official capacity), United Kingdom, Italy, United States, Bulgaria, Czechoslovakia, Poland, Romania, Soviet Union, Brazil, Burma, Ethiopia, India, Mexico, Nigeria, Sweden, and the United Arab Republic (UAR). The ENCD began work in 1962 and, before it was reconstituted in 1979 as the UNDC, met over 430 times~\cite{ENDC_1}. 

The result of those meetings was the NPT, which opened for signature in 1968, and entered into force in 1970.  A total of 191 States have joined the Treaty, including the five nuclear-weapon states recognized by the NPT. More countries have ratified the NPT than any other arms limitation and disarmament agreement, underscoring its global significance.

\subsection{Scope of the NPT}\label{sec:scope}
The concept of \textit{nuclear nonproliferation} is intrinsically tied to the NPT and refers to a specific risk to global security –- the risk of creating nuclear warheads. The three pillars of the NPT are to (1) prevent the spread of nuclear weapons and weapons technology, (2) promote cooperation in the peaceful uses of nuclear energy, and (3) further the goal of achieving nuclear disarmament and general and complete disarmament. ~\cite{NPT}.  

Under the NPT, parties that are non-nuclear-weapon states have committed to not manufacture or otherwise acquire nuclear weapons or other nuclear explosive devices, while parties that are nuclear-weapon states have committed not to in any way assist, encourage, or induce any non-nuclear-weapon state party to manufacture or otherwise acquire nuclear weapons or other nuclear explosive devices.  A ``nuclear-weapon state" party under the NPT means a state that manufactured and exploded a nuclear weapon or other nuclear explosive device before 01 January 1967. The NPT further requires non-nuclear-weapon state parties to  enter into safeguards agreements with the IAEA, to verify compliance under the NPT  (as discussed in more detail in Section~\ref{sec:safeguards})~\cite{NPT}.

With respect to the restrictions applicable to sharing nuclear material, equipment, or underlying technology, the NPT is carefully drafted to establish controls over only ``(a) source or special fissionable material, or (b) equipment or material especially designed or prepared for the processing, use or production of special fissionable material"~\cite{NPT}. The choice of NPT scope was carefully negotiated such that, considering both nuclear weapons and peaceful uses of nuclear technology, controls should be only implemented to ``the extent necessary" to ensure peaceful use.

Source or special fissionable material are specific terms well understood around the world.  Source material means unenriched uranium or thorium, and special fissionable material means material enriched in uranium (U-235), uranium-233, and plutonium-239~\cite{IAEA_glossary_safeguards}.  Uranium, thorium, and plutonium are the large-atomic-mass elements that can themselves undergo (or be bred/configured to undergo) self-sustaining chain reactions: the fissioning of one nucleus leading to fissioning of multiple other nuclei.  These rapidly growing fission chain reactions are the foundation for all nuclear weapons.

The phrase ``equipment especially designed or prepared for the processing, use or production of special fissionable material" is also well-established and specific.  It refers to (i) equipment ``especially designed or prepared for" creating or isolating special fissile materials, that is, the so-called nuclear fuel cycle, such as  nuclear enrichment facilities (which can isolate fissile uranium-235); and (ii) nuclear fission reactors (which inherently process and use special fissionable material, and also create fissile plutonium or uranium-233 from uranium or thorium, even under normal operation).

The phrase ``especially designed or prepared for" plays an important role in clarifying the scope of equipment and technology brought under the nonproliferation regime. Everything from fiber optics to steel to screws can play a role in the construction of a nuclear weapon. But under the NPT, only that equipment ``especially designed or prepared for" playing a role in the direct nuclear fuel cycle is included, allowing the global nonproliferation framework, and the implementing network of domestic law and regulations that implement it, to focus on those elements of key concern.

The long-standing list of key equipment determined to be ``especially designed or prepared for" the production or use of special fissionable material, and thus captured within the nuclear nonproliferation regime, is found in the ``Trigger List" established by the Zangger Committee (also known as the ``NPT Exporters Committee"), a 39-member coalition that is a key interpreter for what constitutes ``especially designed or prepared material" under Article III.2 of the NPT~\cite{NPT,ZanggerCommittee}. This list includes nuclear reactors, fissile fuel fabrication facilities, reprocessing plants, enrichment facilities, heavy water facilities, conversion facilities, and key components and materials~\cite{ZanggerCommittee}. Each of these supports a nuclear end use -- an end use related to processing, using, or producing special fissionable material as required by the NPT. A very similar Trigger List was also established by the Nuclear Suppliers Group (NSG) to also reflect those components ``especially designed or prepared for" nuclear end use, and which should directly be subject to safeguards under the NPT.\footnote{A key example of this distinction lies around the nuclear nonproliferation regime's treatment of deuterium. The Trigger List extends to deuterium specifically for nuclear end use (as heavy water -– water enriched in deuterium -– can be used to breed special fissionable material). However, deuterium not destined for a nuclear reactor falls outside of the Trigger Lists and the NPT.}   ``The guiding question for listing items on the Trigger List is `do the items meet the EDP [especially designed or prepared] criteria for the processing, use, or production of special fissionable material?'" 

The Trigger Lists play a very important role in interpreting the scope of facilities covered under the NPT.  \textit{Critically, they explicitly exclude fusion from the scope of these lists.}  Both state in their note under nuclear reactors, for example, ``this entry does not control fusion reactors."  Both Trigger Lists also exclude tritium and tritium production technologies (although they feature on a separate list focused on dual-use\footnote{\textit{Dual-use} technologies are those technologies that are designed or suitable for both civilian and military, terrorism, or weapons of mass destruction-related applications.} export controls).  

Even as the global nuclear nonproliferation regime has aged, it has remained steadfastly focused on the production chain and fuel cycle for fissile nuclear materials, including as implemented by the IAEA through its safeguards program.

\subsection{IAEA safeguards}\label{sec:safeguards}
The signature feature of the NPT-based nonproliferation regime is safeguards.  In practice, safeguards is monitoring and accounting of source and special fissionable material – taking the form for example of IAEA inspectors routinely using detectors to measure the amount of (fissionable) nuclear material at various declared sites around the world and affirming that the measured quantity matches the declared amount~\cite{Doyle,PANDA,pandendum,ensslin_guide}. The International Atomic Energy Agency (IAEA), a global United Nations body based in Vienna, Austria, is  entrusted with the key verification responsibilities under the NPT.  

Under the NPT, Article III, each non-nuclear-weapon state party is required to conclude a comprehensive safeguards agreement (CSA) with the IAEA to enable the IAEA to verify the fulfillment of their obligation under the NPT, with the goal of preventing diversion of nuclear material from peaceful uses to nuclear weapons or other nuclear explosive devices.  Under a CSA, the IAEA ensures that safeguards are applied on nuclear material in the territory, jurisdiction or control of the state for the purpose of verifying that such material is not diverted to nuclear weapons or other nuclear explosive devices. As of September 2021, 178 non-nuclear-weapon state parties to the NPT have brought into force CSAs required by the NPT~\cite{NPT,IAEA_Legal,IAEA_CSA}.

In addition to a CSA, there are three other types of safeguards agreements.  The five nuclear-weapon states party to the NPT have voluntarily entered into safeguards agreements under which the IAEA applies safeguards to nuclear material in facilities that the state has voluntarily offered and the IAEA has selected for the application of safeguards. Safeguards are also implemented in three countries that are not party to the NPT -- India, Pakistan, and Israel -- on the basis of item-specific agreements they have concluded with the IAEA to ensure safeguarded facilities and materials are not used for nuclear weapons proliferation~\cite{IAEA_CSA,IAEA_China,IAEA_France,IAEA_Russia,IAEA_UK,IAEA_US,IAEA_ItemSpecific}. Additionally, the Small Quantities Protocol (SQP) is a protocol available to states with limited nuclear material and holds in abeyance many procedures in Part II of a CSA. The purpose of an SQP is to reduce safeguards implementation burdens in states with limited nuclear activities, while retaining the integrity of the broader safeguards system~\cite{SQP}. Safeguards generally apply to facilities and activities covered by the Zannger Committee and NSG Trigger Lists (see Annex II of the Model Additional Protocol)~\cite{ZanggerCommittee,AdditionalProtocol}

In 1997, the Model Additional Protocol~\cite{AdditionalProtocol,AdditionalProtocol_NRC} was created to enhance IAEA inspection capabilities following the discovery of clandestine weapons programs in the 1990s.  The Model Additional Protocol extended the scope of the NPT regime to undeclared “nuclear fuel cycle activities” including nuclear fission reactors, as well as broadened the ability of the IAEA to conduct environmental inspections of non-nuclear sites upon IAEA request. However, it still maintains a focus on only applying safeguards to those activities specifically related to the production or possession of special fissionable material, tied to the Trigger Lists discussed above. For example, although tritium (hydrogen-3) by this time was commonplace in nuclear weapons as a booster, the Additional Protocol intentionally did not extend safeguards requirements to tritium. Moreover, it maintained the definition of ``nuclear reactor" to those devices ``capable of operation so as to maintain a controlled self-sustaining fission chain reaction," excluding fusion.\footnote{The long-standing clarifications in the Model Additional Protocol and Trigger Lists as to the definition of nuclear reactor gives significant weight to a conclusion that the definition of ``nuclear reactor" in the IAEA statute also excludes fusion devices (e.g., as used in Article XII, concerning safeguards).} This long-standing focus on source and special fissionable materials and fission process, even in updates to the global regime, reflects the NPT’s intent to control the use of atomic energy specifically “to the extent necessary” to ensure its peaceful use.  

\subsection{Export control framework}\label{sec:export_controls}
International trade is subject to a network of export controls to protect each country's national security interests, promote foreign policy objectives, and leverage economic interests and opportunities. There is a robust framework of  export control regimes to prevent the proliferation of weapons of mass destruction and prevent destabilizing accumulations of conventional weapons and related material. Some of this framework is derived from international nonproliferation obligations under the NPT, while other elements of this framework (particularly the \textit{dual-use} export controls framework) more broadly police other international and domestic obligations and agendas.

\subsubsection{Nuclear-specific export controls implementing the NPT}
For example, after IAEA safeguards, countries implement their NPT treaty obligations into their domestic law.  For example, in implementing the NPT in the US, Congress among other things passed the Nuclear Non-Proliferation Act of 1978~\cite{PubLaw_Nonpro} and Atomic Energy Act of 1954~\cite{PubLaw_AEA}, which gave the US Nuclear Regulatory Commission and US Department of Energy stronger authority to regulate exports of source and special nuclear material (the US analog of ``special fissionable material") and those technologies ``especially designed or prepared for" producing or using special nuclear materials -- i.e., those items and technologies falling under the NPT. See Refs.~\cite{EC_NRC1} and~\cite{EC_NRC2} explaining how the Nuclear Non-Proliferation Act gave the NRC jurisdiction over exports of ``nuclear facility components and other nuclear items and substances having possible significance for nuclear explosive uses" in 10 CFR Part 110~\cite{Part_110}, and Ref.~\cite{EC_NRC3} explaining how the Atomic Energy Act and Nuclear Non-Proliferation Act give the Department of Energy jurisdiction over exports of nuclear technology that poses a nuclear proliferation hazard because it can produce special nuclear material in 10 CFR Part 810~\cite{Part_810}. These are generally seen as strict export controls, and for example typically require license or notification for \textit{any} export, regardless of destination or use. The technologies and materials captured under these nuclear-specific export controls tie to the Trigger List, and a general requirement for export licensing is that the destination applies safeguards to this regime, including appropriate reporting to the IAEA under the Additional Protocol.

\subsubsection{Dual-use export controls}
Beyond nuclear-specific export controls that directly implement the NPT, a broader level of export controls exists across multiple countries that police exports of technology, material, or ancillary equipment including when they have a dual use, which could tangentially relate to nuclear end use and proliferation. Specifically, ``dual use" refers to items that have both commercial and some military or proliferation applications, but are not ``especially designed or prepared" for nuclear end use.\footnote{As explained by the IAEA, in the nuclear context ``dual use items are items that can be used either for a nuclear or for another purpose, or for a variety of non-nuclear purposes, and hence do not fall within the scope of Article III.2 of the NPT which relates only to `equipment or material \textit{especially designed or prepared} for the processing, use or production of special fissionable material' [emphasis added]."}

This division between application of NPT safeguards and nuclear-specific export controls on source and special fissionable material and the nuclear fuel cycle (captured in the Trigger Lists), and export controls for dual-use technology that may have a nuclear misuse concern, is not simply happenstance but is well understood.  The concept of applying dual-use export controls to non-NPT dual-use technologies and materials appeared early on, with the NSG issuance of additional guidance separate and beyond the Trigger List on dual-use exports (the NSG Part 2 Guidelines) in INFCIRC/254 in 1992. Despite formally recognizing in the early 1990s that there are technologies that if misused could have an impact on proliferation, the Additional Protocol was not expanded to capture those technologies under safeguards.  The IAEA instead appeared content on letting such technology be managed by export controls.  One of the key goals of the dual-use export control regime is thus to ensure that there is not diversion of the technology to nuclear weapon applications.   Although discussed more below, fusion related technologies such as tritium development systems are already captured under this regime.

In the US, these broader export controls are found in the Department of Commerce's Export Administration Regulations, EAR, found at 15 CFR Part 730-744~\cite{Part_730}. These regulations (save certain limited export reporting obligations) largely find their authority \textit{not} in the NPT, but in domestic laws, such as in the US the Export Administration Act of 1979~\cite{PubLaw_EC_1979} (as updated by the Export Control Reform Act of 2018~\cite{PubLaw_EC_2018}).

The Department of Commerce dual-use export controls (15 CFR Parts 730-744) sit outside the purview of the NPT and what is commonly understood as the nuclear nonproliferation regime (although there is some coordination~\cite{EC_NRC1}), and cover beyond what is ``to the extent necessary" for the NPT to cover under safeguards. For example, while the NRC and US Department of Energy (DOE) regulate the export of nuclear fission reactors under 10 CFR 110 and 810 (as they process and produce special nuclear material), the Department of Commerce regulates the exports of equipment apart from the central nuclear island at a nuclear fission plant (e.g., ECCN 2A2901 and 2A291~\cite{EC_DOC}), such as radiation monitors, simulators, and backup diesel generators. Such items or ``balance of plant" components are those that help a nuclear power plant run but are not themselves part of the nuclear fuel cycle. 

Said another way, the nuclear-specific export controls by the NRC and DOE exist when the item being exported has direct significance to nuclear proliferation (like the nuclear fuel cycle), while the Department of Commerce regime covers those exports of technologies that themselves have limited risk, but if abused or misused could create a proliferation concern. This distinction between the role of the NRC and DOE as to policing core proliferation concerns, versus the Department of Commerce as a backstop dual use regulator, is seen most clearly in a recent action by the NRC to clarify that the NRC export control regulations did not  apply to deuterium exported for non-nuclear end uses. This decision was made on the grounds ``that deuterium for non-nuclear end use is not an item or substance that is `especially relevant from the standpoint of [the Nuclear Non-Proliferation Act] export control because of [its] significance for nuclear explosive purposes'" and therefore better suited for control by the US Department of Commerce~\cite{FR_Deuterium}. 

The Department of Commerce export regime explicitly balances economic and national security interests (see Section 2 of the Export Administration Act of 1979~\cite{PubLaw_EC_1979}). Thus, dual-use export controls may touch on a number of ancillary technologies that may -- if misused -- have an impact on proliferation, but have a much more proportionate hand.  For example, exports of nuclear balance of plant components only as a general rule need a license when exported to a specific set of countries, with some exceptions (See the Commerce Country Chart, Ref.~\cite{Part_738}). Tighter controls can exist for certain equipment or materials, but licensing can often be prompt, and there is more regulatory flexibility. That being said, the regime is still quite strict. As described in the NSG Guidelines, countries should not permit exports of dual use items:
\begin{itemize}
	\item for use in a non-nuclear-weapon state in a nuclear explosive activity or an unsafeguarded nuclear fuel-cycle activity, 
	\item in general, when there is an unacceptable risk of diversion to such an activity, or when the transfers are contrary to the objective of averting the proliferation of nuclear weapons, or
	\item when there is an unacceptable risk of diversion to acts of nuclear terrorism.
\end{itemize}

\subsubsection{Summary of the export control framework}
\begin{figure}[H]
	\centering
	\includegraphics[width=\linewidth]{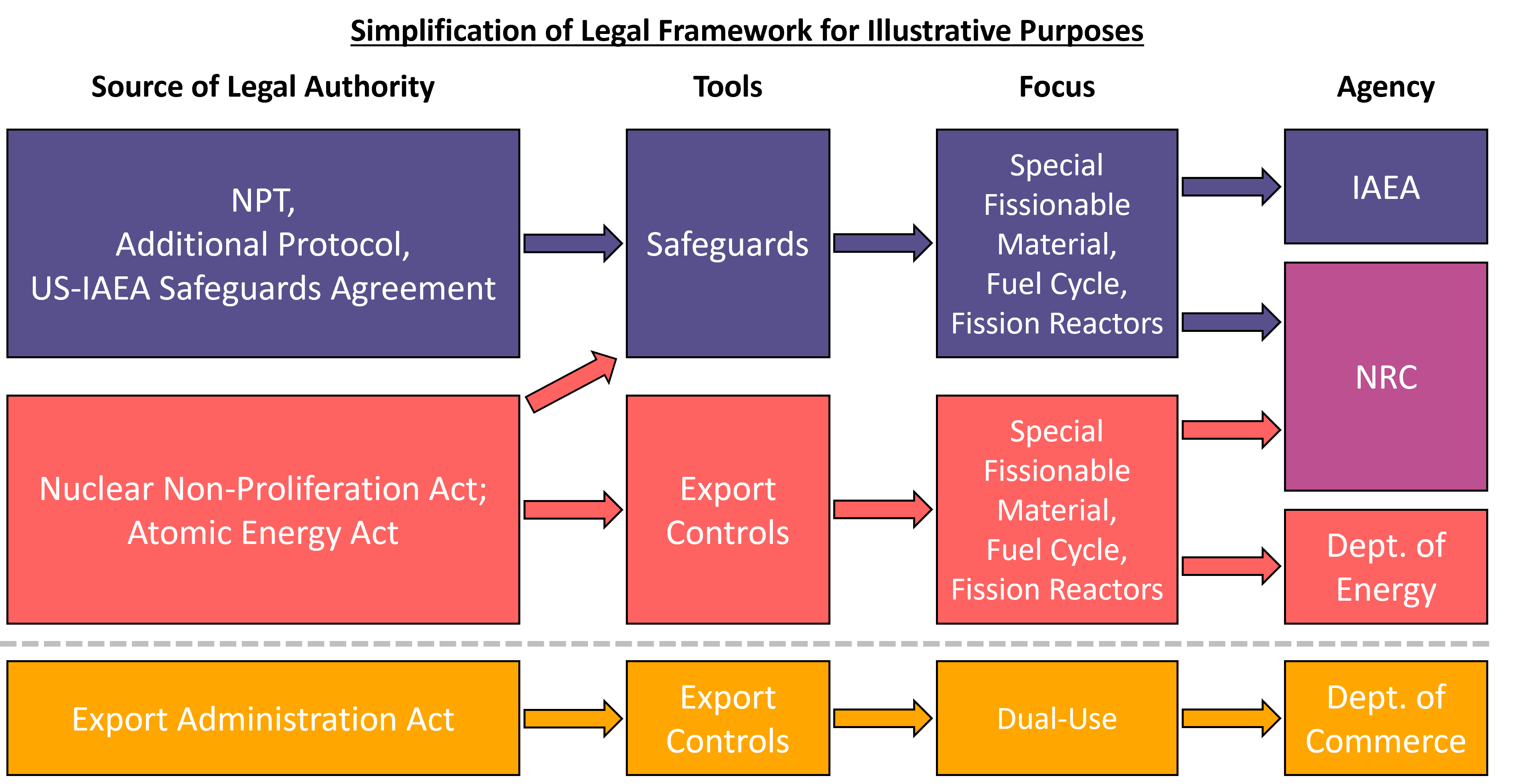}
	\caption{Simplified overview schematic of the security and control agencies, tools, focus of the tools, and sources of legal authority specific to the US as discussed in the legal background section. The portion above the dashed line focuses on special fissionable material, fuel cycle and nuclear fission reactors, whereas the portion below the line is ``non-nuclear".}
	\label{fig:legal_overview}
\end{figure}

A schematic of the agencies, tools (i.e., safeguards and export controls), focus of the tools, and connection to treaties and acts discussed in this section is shown in Figure~\ref{fig:legal_overview}. 
\subsection{Applying fusion to the NPT, IAEA safeguards, and export control framework}\label{sec:fusion_spec}
On its face, fusion does not fall under the NPT.  As noted, the plain language of the NPT is expressly limited in scope to source or special fissionable material, or such related ``equipment" that is ``especially designed or prepared for processing, use or production of special fissionable material." The NPT regulates source material because it can be enriched into special fissionable material; it regulates special fissionable material because it is or can be enriched into material for use in nuclear weapons; and it regulates certain specific underlying equipment because such equipment ``is especially designed or prepared" to make, use, or further refine special fissionable material.  NPT implementing documents, set forth in IAEA safeguards agreements and protocols to the agreement are likewise limited to this same type of equipment and material.  For example, fusion is not included in the safeguards agreements, Additional Protocol, or the Trigger List. 

A review of the history of the NPT also confirms that not including fusion appears to be intentional. To this point, the US Permanent Representative to the United Nations said on 15 May 1968 regarding the NPT that ``controlled thermonuclear fusion technology will not be affected by the treaty"~\cite{NPT_noFusion1,NPT_noFusion2}, an intent that was recognized and mirrored by the depository notes provided by other nations. For example, Germany again quotes that its interpretation of the NPT was that ``controlled thermonuclear fusion technology will not be affected by the Treaty,"~\cite{NPT_Germany,NPT_Germany2} and Japan considered that ``thermonuclear fusion reactors should not come under the prohibitions of the NPT"~\cite{NPT_Book}. 

Current US treatment of fusion technologies echoes this categorization of fusion outside of the NPT. The US implementation of the NPT takes the form of safeguards that applies to nuclear reactors and nuclear materials (see, for example, the NRC's regulations in 10 CFR Part 74 that set forth materials control and accounting requirements for special nuclear material, those in 10 CFR Part 73 that set forth physical security requirements~\cite{Part_74,Part_73}, and those in 10 CFR Part 75 that apply additional IAEA safeguards requirements, none of which apply to fusion).

Even if outside the NPT, under international trade laws, the existing export control framework still applies to fusion. And this is one familiar with fusion, especially with the development of the ITER fusion facility, which is a 35-country fusion research and development project in Southern France. Not only is the equipment and underlying technology exporting to France, but each ITER collaborator is entitled to access each other’s technologies and export controls come into play. 



\section{Evaluation of fusion against proliferation risks}\label{sec:meat}
Given that fusion currently falls outside of the NPT  regime and safeguards, two options exist for evaluation.  First is the active addition of fusion into the NPT regime.  Given the long-standing exclusion, this would require at a minimum modification of various international agreements if not the NPT itself.  An alternative is to turn to the dual-use export control regime that currently polices misuse of technologies that could, if misused, produce special fissionable material or support weapons production.  Adopting fusion into this regime (which in large part has already occurred), would require less fundamental change to the overall nonproliferation framework.  

\begin{figure}[H]
	\centering
	\includegraphics[width=\linewidth]{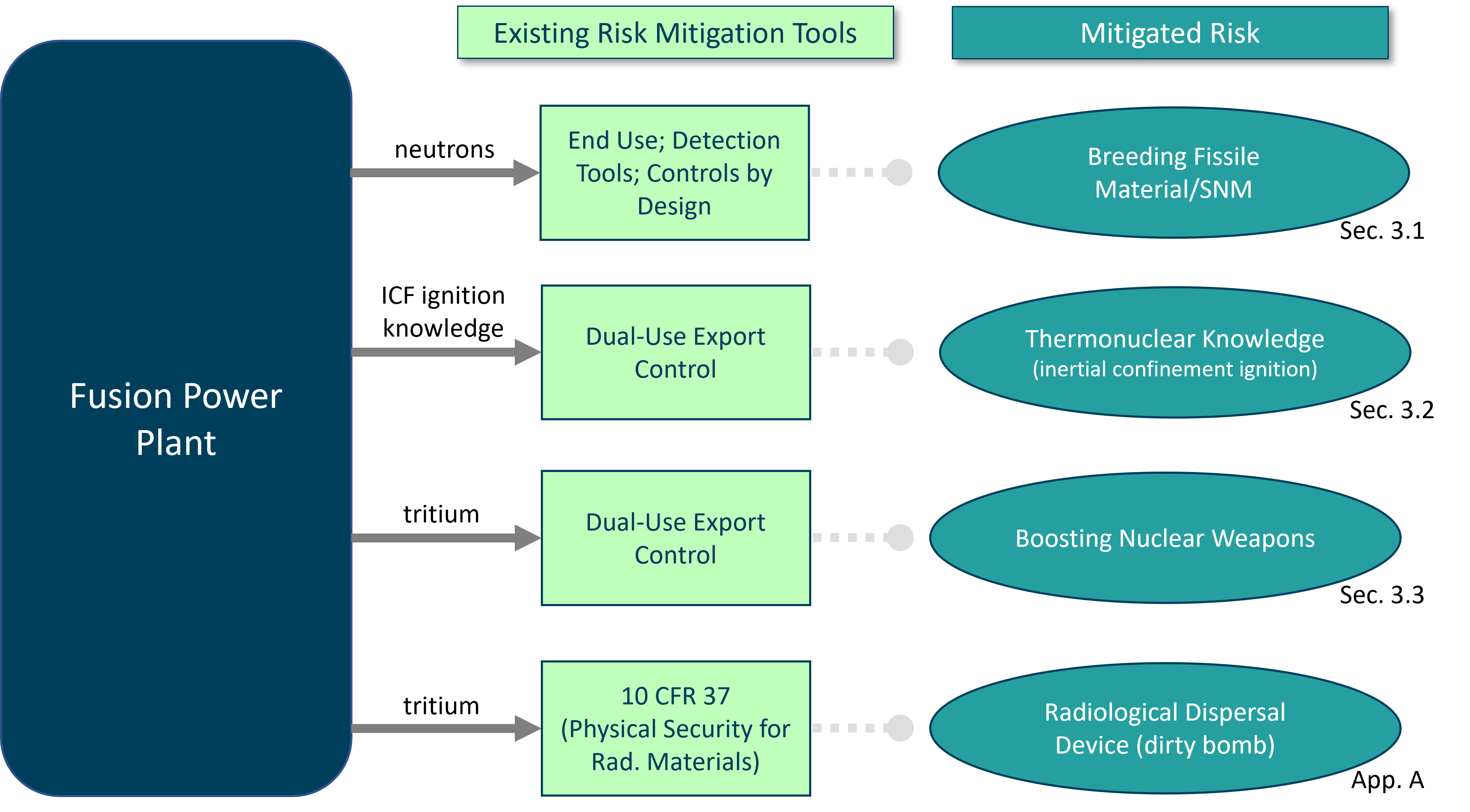}
	\caption{Overview of the potential proliferation risks associated with fusion power plants and the already existing tools that can be used to mitigate the risks.}
	\label{fig:overview}
\end{figure}

The technical case plays a key role in choosing between these options.  In order to evaluate these options, this section evaluates the theoretical nonproliferation risks that have been identified related to fusion: use of neutrons to create fissile material (see Section~\ref{sec:SNM}), leveraging research of ignited, inertially confined plasmas for thermonuclear weapon development (see Section~\ref{sec:knowledge}), and diversion of tritium to boost nuclear weapons (see Section~\ref{sec:nukes}). Note that the ancillary risk of diversion of material towards a radiological dispersion device is separately addressed (see Appendix~\ref{sec:RDD}). The risks are evaluated in light of whether they create an inherent and immediate special fissionable material proliferation risk that warrants safeguards, or a more indirect risk of abuse that is appropriately addressed via dual use export controls. Figure~\ref{fig:overview} summarizes these risks and the corresponding, existing mitigation measures discussed in this section. 

\subsection{Use of neutrons to breed special fissionable material}\label{sec:SNM}
Fusion power plants can produce substantial neutron fluxes, and these neutrons can theoretically be used to breed significant quantities of special fissionable material (e.g., plutonium, enriched uranium, and U-233) from source material (e.g., depleted/natural uranium and thorium). The IAEA has defined significant quantities as ``the approximate amount of nuclear material for which the possibility of manufacturing a nuclear explosive device cannot be excluded"~\cite{IAEA_glossary_safeguards}. The significant quantity for plutonium (containing less than 80\% Pu-238) or U-233 is 8 kg, and for highly enriched uranium (greater than 20\% U-235) it is 25 kg. Significant quantities are also defined for indirect use nuclear material and are 20 ton for thorium, 10 ton for natural uranium, 20 ton for depleted uranium, and 75 kg of uranium-235 in any uranium that is less than 20\% U-235~\cite{IAEA_glossary_safeguards}.

Prior works have investigated fission-fusion hybrid systems designed to breed fuel for fission reactors from fusion neutrons~\cite{fusion_breeder}. Although fission-fusion hybrids are beyond the scope of this paper (and already receive distinct treatment because of the inherent fission from the outset), the idea that fusion power plants could be abused to breed special fissionable material is considered. Prior works have investigated this misuse, notably Refs.~\cite{goldston, glaser}, and their work is summarized and further discussed in this section. Three scenarios are analyzed, which are considered comprehensive:
\begin{enumerate}[A.]
	\item Clandestine production of special fissionable material (action at an unknown facility),
	\item Covert production of special fissionable material (unknown action at a known facility), and
	\item Known production of special fissionable material (known action at a known facility, akin to a so-called ``breakout scenario" where a state openly uses the technology to produce special fissionable material).
\end{enumerate}

\subsubsection{Clandestine production: action at unknown facility}
Clandestine production of special fissionable material requires an unknown facility, tons of source material, and a sufficiently small footprint to avoid detection -- if detected, this clandestine production either stops, or becomes the known production scenario discussed further below. Thus, the feasibility of an unknown, undeclared facility is examined.

A `smallest device possible' case is considered (because smaller devices are more likely to avoid detection). Studies such as those by Kuteev et al. have been conducted to determine the minimum size fusion device that can emulate the production of fusion neutrons (see Ref.~\cite{fissionFusionHybrid} and other concepts e.g., in Ref.~\cite{fissionFusionHybrid2}). Based on optimistic extrapolations of plasma physics, device engineering, and fusion technology according to Ref.~\cite{glaser}, a theoretical device could draw 40 MW of continuous grid power to produce 1.8 MW of continuous fusion power. Assuming an 85\% duty factor and that 80\% of neutrons are captured in a breeding blanket (containing uranium or thorium), this device could hypothetically produce 3.5 kg of plutonium or U-233 per year~\cite{glaser}. This production rate is consistent with legacy fission-fusion hybrid analyses that estimate the production of 0.64 plutonium or U-233 atoms per 14.1 MeV neutron (from deuterium-tritium fusion), or 2.85 kg of plutonium per MW-year of DT fusion power generation. Note that lower energy neutrons (e.g., 2.45 MeV neutrons from deuterium-deuterium fusion) create fewer plutonium and U-233 atoms since, for one, there is less neutron multiplication.

A facility housing and operating such a device would be easily visible due to the required subsystems. Reference~\cite{glaser} notes that the 40 MW of continuous input power and the necessary cooling require large electrical supply lines, large power conversion buildings, and a significant cooling facility. Additionally, the device would require a substantial shielding structure, a remote operating facility, and, in the case of the concept proposed in Ref.~\cite{fissionFusionHybrid}, large neutral beam injection systems that are used to heat the plasma.

A facility housing and operating such a device would also have detectable radiological signatures in the environment, for which monitoring tools exist under the IAEA Additional Protocol that the international community can leverage. Sample environmental monitoring data from the Tokamak Fusion Test Reactor (TFTR) and National Spherical Torus Experiment (NSTX) at the Princeton Plasma Physics Laboratory are shown in Tab.~\ref{tab:PPPL}; TFTR operated from 1995-1997, 1998 was a transitional year from TFTR to NSTX, NSTX operated from 1999-2008, and TFTR underwent decommissioning activities between 1999-2002~\cite{PPPL_1995,PPPL_1996,PPPL_1998,PPPL_1999,PPPL_2000,PPPL_2001,PPPL_2002,PPPL_2004,PPPL_2005}.  Note that data are skewed and do not correct for rain and groundwater interactions, nor do they include the distinction between elemental tritium (HT) and tritiated water (HTO) releases. Further note that the levels are much lower than the US Environmental Protection Agency's tritium concentration limit for drinking water (20,000 pCi/L)~\cite{Part_141}. The data do demonstrate that even trace levels of tritium lost from the facility are detectable in environmental sampling, particularly the trends from 1995-1997. Beyond tritium sampling, the fertile (uranium and thorium) and fissile material leave environmental signatures.

\begin{table}[H]
	\centering
	\caption{Data from annual site environment reports at the Plasma Princeton Physics Laboratory for the TFTR and NSTX experiments. The effective dose equivalent (EDE) is at the site boundary and due only to tritium, onsite surface means maximum tritium concentration in onsite surface water, and onsite soil means maximum tritium concentration in offsite soil samples.}
	\begin{tabular}{rrrrr}
		\toprule
		\multicolumn{1}{l}{Year} & \multicolumn{1}{l}{Airborne T Released [Ci]} & \multicolumn{1}{l}{EDE [mrem]} & \multicolumn{1}{l}{Onsite Surface [pCi/L]} & \multicolumn{1}{l}{Offsite Soil [pCi/L]} \\
		\midrule
		1995  & -     & 0.097 & 789   & 790 \\
		1996  & -     & 0.320 & 1288  & 2845 \\
		1997  & -     & 0.430 & 2472  & - \\
		1998  & 592   & 0.120 & 1441  & - \\
		1999  & 348   & 0.156 & 1041  & - \\
		2000  & 728   & 0.211 & 1104  & - \\
		2001  & 184   & 0.618 & 3109  & - \\
		2002  & 2090  & 0.251 & 479   & - \\
		2003  & 451   & 0.089 & 225   & - \\
		2004  & 353   & 0.057 & 502   & - \\
		2005  & 219   & 0.011 & 365   & - \\
		2006  & 514   & 0.014 & 236   & - \\
		2007  & 234   & 0.016 & 257   & - \\
		2008  & 171   & 0.007 & 5473  & - \\
		\bottomrule
		\bottomrule
	\end{tabular}%
	\label{tab:PPPL}%
\end{table}%

Lastly, the fertile material, which is itself controlled, would need to be covertly diverted from a safeguarded nuclear fuel cycle or itself be clandestinely produced. Taken together, clandestine production of special fissionable material is not credible since the source material is controlled, the holistic footprint would be readily visible, and the production would inherently leave detectable environmental signatures.

\subsubsection{Covert production: unknown action at known facility}
Covert production, where nefarious activity is performed at a known facility, is not restricted to the small-as-possible fusion device of the previous subsection on clandestine production since the facility need not be hidden. Thus, Glaser and Goldston (Ref.~\cite{glaser}) consider a 2500 MWth, 2660 MW plasma power device with an incident neutron rate of $9.42\times10^{20}$ n/s. For simplicity, this paper will consider the same device; however, it is noted that (i) there is a breadth of commercial fusion concepts that range from a few MW to a few GW~\cite{FIA_survey_2021,FIA_survey_2022}, (ii) there are different energy recovery concepts that change the fusion reactions per MWe, and (iii) there are different fuel types that change neutrons per MW and neutron energy. All of these considerations will ultimately change hypothetical production rates of special fissionable material in a reconfigured device.

Before discussing the theoretical production rates of plutonium and U-233, it is important to note that the quantity of fertile material injected would need to be on the order of 500 tons to maximize fissile-material production. Thus, the quantity of fertile material may already constitute a significant quantity. This fertile material, especially at this order of magnitude, is also already controlled.

Assuming the fusion device and quantity of fertile material above, Glaser and Goldston estimate that a maximum of 20 kg of plutonium or U-233 could be produced per week (2.5 significant quantities). The limiting factor for plutonium production is heat loading considerations. The limiting factor for U-233 production is running out of fuel, i.e., not consuming more tritium than is produced (since neutrons captured in thorium are hence not being captured to produced tritium). If the device does not require a closed fuel cycle due to other sources of tritium, the U-233 production rate could theoretically be increased to 28.85 kg per week.

Covert operation in a known facility, such as a regulated facility in the United States, would be practically impossible. The process of breeding special fissionable material entails introducing and exposing source material (itself controlled material) to fusion neutrons and then extracting the material after irradiation (e.g., reprocessing and enrichment). While source material could be introduced in the coolant or tritium-producing lithium blanket, detecting these materials is straightforward with existing capabilities; the detection timescales are on the order of seconds to minutes for direct measurement or inspection. These capabilities include:
\begin{itemize}
	\item Easy visual observation of dramatic device changes.
	\item Detection from general environmental sampling for radiological signatures of fertile and fissile material since fusion devices do not require uranium\footnote{Some fusion facilities may use uranium getterbeds to store tritium. However, these getterbeds would generally not be located close to the device in a large neutron flux. In addition, uranium getterbeds are not required and titanium, zirconium-iron, or other materials can be used instead.} nor thorium and should not possess these near the device (let alone hundreds of tons); and, 
	\item Simple tools like portal monitors and metal detectors (currently used at border crossings and airports, for example) at the entrance and exit that can be used to detect source or special fissionable material entering or exiting these facilities.
\end{itemize}

Thus, covert production of special fissionable material at a fusion power plant is not a credible threat. These tools can be captured in the licensing basis of US facilities, and indeed facilities are likely to have pertinent detectors anyway to ensure proper device operation. These tools can also be requirements to meet export control approval thresholds -- export controls are discussed in greater detail in Section~\ref{sec:export_controls} and~\ref{sec:breakout}. Physical security protocols of the type envisioned in the NRC's 10 CFR Part 73 safeguards regime are not applicable here, as those protocols are designed around protecting a significant quantity of special nuclear material that already exists. For fusion power plants, there should be no such material and as soon as \textit{any} is detected, authorities would be alerted. Once authorities are alerted, this covert production scenario if continued becomes the known production scenario discussed in Section~\ref{sec:breakout}. 

\subsubsection{Known production: known action at known facility}\label{sec:breakout}
This final scenario is breakout, wherein a state openly abuses a fusion power plant to produce special fissionable material. The timescale to produce relevant quantities of material depends on the power density of the fusion device and can range from months to years. It is important to recognize that a core focus of the nonproliferation regime is on \textit{detecting} improper use of nuclear materials -- it is not designed to solve a breakout scenario that occurs. Indeed, the early detection of a breakout scenario is itself a success of this regime, as it enables states to then bring multiple tools to bear to isolate and coerce a bad actor into compliance. 

Although solving a breakout scenario is beyond the scope of the current nonproliferation requirements, it is worth noting that fusion offers many benefits that aid in resolving such scenarios. First and foremost, a clarion distinction between breakout at a fission power plant and breakout at a fusion plant is that the fission plant already contains uranium, plutonium, and spent nuclear fuel. This characteristic of fission plants raises an availability challenge and also a policy challenge -- ``as was seen in North Korea, little can be done short of invasion to prevent a determined sovereign nation from processing its own Pu for use in nuclear weapons. Massive bombing of the spent fuel in Yongbyon would have resulted in widespread contamination, an unacceptable consequence. More limited bombing would have required frequent repetition to prevent access to the fuel buried in the rubble -- also an unacceptable consequence"~\cite{goldston}. In stark contrast, a fusion facility does not contain fissile material at the time of breakout and options exist to prevent operation without risk of contamination. In addition, most commercial-scale fusion concepts have advertised not requiring post-shutdown cooling~\cite{NRC_FIA}, whereas most fission facilities do require such cooling to avoid disaster. Thus, fusion power plants can generally be immediately shut down with few repercussions and many subsystems, such as power electronics and magnet cooling, can be disabled with minimal risk of contamination.

To the extent that limited risks may exist in regard to improper use of a fusion device, it is in the form of \textit{improper modification} of commercial technology and devices. A commercial fusion device would not be designed to physically accommodate a special fissionable material program. The process of breeding special fissionable material entails introducing and exposing source material to fusion neutrons in a specific manner, and then extracting and purifying the material after irradiation. These facilities do not exist in a commercial fusion device, and the existing equipment (e.g., pumps, tritium-producing blanket) would require substantial modification. Such modifications would likely have deleterious effects on the fusion device operation, particularly on the fuel cycle in a way that would require additional, external fuel sources. Thus, the risks above extend to devices that must be especially designed or substantially modified to be misused. In fact, Glaser and Goldston note that a 1-2 month period could produce one significant quantity of special fissionable material and that the majority of this time is dominated by the time required to reconfigure and restart the facility~\cite{glaser}. This is a critical distinction from fission power, where operation of the fission reactor \textit{as intended} creates special fissionable material that can be used to create a nuclear warhead.

This lesser risk for fusion is thus fundamentally addressable and addressed by the dual-use export control regime that applies to everything from computer chips to high-end metals -- items that in themselves are not dangerous unless modified and/or misused by an improper actor (the concern Congress outlined in Ref.~\cite{EC_NRC1}). For all scenarios above and the two sections that follow (including thermonuclear knowledge and tritium), the existing export control regime provides protection against the transfer of relevant technology (along with many other non-fusion technologies) to bad actors -- particularly those known to be at risk of diverting technology across the spectrum for improper purposes. In the US for example, Supplement 1 to 10 CFR Part 740 of the US Export Administration Regulations identifies a small selection of approximately 10 destinations (particularly the Country Group D:2 ``nuclear" destinations) that have already been evaluated as high risk from a diversion perspective, and for which exports of fusion technology can be restricted. Not much would need to change in practice.  As discussed above, the NSG already envisioned application of dual-use export control to such fusion technologies, including certain neutron generators, key technologies and materials, and tritium breeding systems.  Fusion technology as a category need not be subject to a categorical export control, while those components that may be prone to abuse would remain incorporated in the dual-use export control regime.  See 15 CFR 742.3(a)~\cite{Part_742}.

The already robust nonproliferation case for fusion can be potentially enhanced by exploring defense-in-depth via a controls-by-design approach. Fusion devices -- particularly those geared for global deployment -- can be engineered at the outset to include controls to (i) enable remote shutoff, (ii) force obsolescence without routine maintenance or component replacement on time scales shorter than those to produce a significant quantity of special fissionable material (e.g., copy-resistant semiconductors that inherently fail after some amount of neutron irradiation), and/or (iii) force digital planned obsolescence. These are technologies in use today, as demonstrated by the simple example of remote lockout of John Deere tractors pilfered by Russian forces in Ukraine~\cite{CNN_tractors}. Ideally, as fusion devices implement a controls-by-design approach, the number of destinations that require a license can be reduced, potentially eventually reaching ``EAR99" level of control.

Additionally, longer-term usage-based controls could be implemented to secure fusion devices from misuse through, for example, a sensor network that evaluates (a variety of information carriers including) the presence of special fissionable material and deactivates the machine upon detection. Usage-based monitoring and controls have been well analyzed. For example, an analysis by Microsoft and OpenAI explains how ``foundational technologies are also best regulated by restrictions on the end users that can have access to them, and the end uses to which they can be put"~\cite{Craig}. This analysis includes user and use verification control, tamper-proof hardware and secure infrastructure, and AI-enabled identification of problematic uses and users~\cite{Craig}. Limited reform of the existing export control regime can incentivize companies to implement these enhanced protections or usage-based controls in order to reduce or eliminate certain licensing requirements. Further research and discussion can be conducted in this area to determine how such a regime could enhance global security while still enabling prompt deployment of fusion once it is commercialized in the near term.

Lastly, the Additional Protocol's enhanced tools such as broad environmental sampling and related inspection of non-nuclear sites and activities can be leveraged to mitigate any concerns about fusion systems without broadly incorporating fusion writ large into the nuclear fuel cycle.  As discussed above, use of fusion to breed special fissionable material requires source material (which is controlled under the NPT), and would be easy to detect given that a fusion device should have no source or special fissionable material at all within the facility.  Thus, the same way that current Additional Protocol's broad sampling tools could be used to enable verification that covert or undeclared industrial or research facilities are not being abused to harbor a covert program, these tools can be applied to help ensure fusion devices are not being so abused.  

Taken together, scenarios involving covert and clandestine abuse of fusion facilities to produce fissile and/or special fissionable material are not credible.  Breakout scenarios nominally are credible, but (i) are themselves conditional signs of success of the nonproliferation regime to prevent covert actions, (ii) are less difficult to address for fusion facilities than for fission reactors, and (iii) can be prevented through application of existing dual-use export controls and in the future a controls-by-design concept.

\subsection{Transfer of thermonuclear knowledge}\label{sec:knowledge}
Inertial confinement fusion (ICF) concepts (but not magnetic confinement fusion concepts\footnote{Plasmas are naturally unstable and must be confined. The amount of fusion in the plasma scales as the product of density and confinement time. Magnetic confinement fusion uses magnetic fields to confine a plasma for long periods of time, whereas inertial confinement fusion compresses a plasma to high densities for a short period of time.}) strive to maximize fusion ignition to achieve very high fusion gains. Some of the knowledge associated with achieving such high fusion gains may be transferable to thermonuclear, two-stage (both fission and fusion) weapon development. In such weapons, the fission core ignites\footnote{Ignition is the fusion operating regime where the energetic, charged particle products are trapped in the plasma long enough to maintain the plasma's temperature at fusion conditions without further, external input energy.} and compresses a deuterium-tritium plasma to fusion conditions (including a very high density), and the core's inertia confines the fusing plasma such that large yields are achieved. This similarity is limited to only those concepts based on ICF using ignited plasmas and particularly the subset using indirect-drive targets as opposed to direct-drive targets~\cite{NAS_ICF}.
 
Note that a commercial fusion facility or net-electricity device is \textit{not} required to perform ICF studies. Current research is performed around the world at facilities like the National Ignition Facility and the Z-Machine at Sandia National Laboratory~\cite{NIF,ZMachine_1,ZMachine_2,ZMachine_3}. In fact, A US National Academy of Sciences report concludes that such research laboratories likely pose greater risk than a commercial power plant~\cite{NAS_ICF}. US laboratory research on ICF has been largely funded by the nuclear weapons program because valuable information can be obtained from ICF experiments that can otherwise only be learned from nuclear testing. A moratorium on such testing by the US was announced in 1992 by President George H.W. Bush, extended by the Clinton administration, and remains in effect. The moratorium was reinforced in 1996 when the US signed the Comprehensive Nuclear Test Ban Treaty (CTBT, though it has not been ratified by the US Senate). The Clinton administration took the position that ICF is not a prohibited activity under the CTBT, and the NPT does allow for laser fusion experiments based on an unopposed, US unilateral statement at the 1975 NPT Review Conference~\cite{NAS_ICF,ICF_NPT_Lasers,ICF_CTBT_Clinton}.

Initially (1960s), all US ICF research was classified. Declassification began in the 1970s; for example, essentially all work on direct drive ICF was declassified by 1974. By the mid 1990s, nearly all information on laboratory ICF experiments was declassified~\cite{NAS_ICF,ICF_declassification}. The notable exceptions are some aspects of computer codes, certain target designs, and some historical experiments related to ICF (particularly the Centurion-Halite program). As computational power increases, and as plasma physics knowledge accumulates in accessible literature, many proliferation concerns associated with ICF become less relevant. Under these circumstances, computer codes would be classified based on their direct application to and/or calibration from nuclear weapons, not according to the physics modeled by the code~\cite{NAS_ICF}. Thus, the significance of thermonuclear knowledge to nuclear weapons proliferation is diminishing. 
 
It is important to recognize as well that the risk of thermonuclear knowledge transfer does not involve the production or transfer of any materials, but instead the transfer of knowledge to bad actors, generally abroad in rogue states. Thus, the \textit{traditional safeguards regime is less relevant} to addressing this concern and in its stead exists an already-robust export controls regime that polices such exchanges. A number of export controls already apply to thermonuclear knowledge, such as the following: 
\begin{itemize}
	\item ECCN 0D999, which includes software for neutron calculation/modeling, software for radiation transport calculations/modeling, and software for hydrodynamic calculations/modeling. 
	\item The general end-use prohibition in 15 CFR 744.2, which prohibits any direct or indirect export of technology that facilitates nuclear weapons development, including research and design.
\end{itemize}
For example, export controls are applied to simulation codes like the Monte Carlo N-Particle (MCNP) code developed by Los Alamos National Laboratory~\cite{MCNP6}. Taken together, fusion power plants do not pose a novel risk regarding thermonuclear weapons physics and the risk posed by a subset of concepts can be treated with already existing export controls.

\subsection{Use of tritium in nuclear weapons}\label{sec:nukes}
Nuclear weapons use neutron chain reactions to quickly induce fission in special fissionable material and generate large amounts of energy~\cite{LANL_primer}. The primary element of all nuclear weapons -- even advanced thermonuclear ``hydrogen bombs" -- is the fission core, as it is the only material capable of self-inducing and sustaining a chain reaction. Tritium and other neutron-rich elements, however, can be added to nuclear weapons to substantially boost their performance. In such ``boosted" devices, the fission component acts as an initiator to achieve and inertially confine a burning plasma wherein deuterium and tritium fuse. This fusion creates more energy and 14.1-MeV neutrons that are themselves damaging or that can induce further fissions in uranium-238, for example. Thermonuclear weapons rely on this fission-fusion-fission cycle~\cite{LANL_primer,glasstone,krane}.

Therefore, tritium has been discussed as a potential proliferation concern because of its use as a booster. However, tritium is only seen as a vertical risk, in that it may enhance a nuclear weapon already in existence -- it does not enable a nuclear weapon to come into being\footnote{Horizontal proliferation refers to a state obtaining nuclear weapon capability, whereas vertical proliferation refers to the advancement of an existing capability.}. The nuclear nonproliferation regime has consistently declined to apply safeguards to vertical proliferation concerns, strongly indicating a lower level of concern with these technologies and practical issues with applying strict NPT-style controls to a vague and expansive group of technologies.  The underlying treaties declined to extend to such components, including the Additional Protocol in the 1990s declining to add tritium as a concern, creating a substantial burden to now add tritium~\cite{AdditionalProtocol}.  As well, nations have licensed facilities that produce substantial quantities of tritium without raising proliferation challenges.  

A key driver for this lower level of concern is that vertical proliferation is only a practical concern for those countries that have already mastered nuclear weapons technology and, in such cases, access to booster materials is not a challenge.  The US Department of Energy published ``the amount of tritium in a [warhead] reservoir is typically less than 20 g"~\cite{DOE_tritium}, and ``if we consider the neutron economy of a fission reactor used to produce Pu [plutonium] and T [tritium],... any nation that is producing Pu for weapons use can, with relatively minor perturbation, produce adequate T to ``boost" these weapons, even taking into account the $\approx$5\% loss of T per year through radioactive decay"~\cite{goldston,glaser}. For example, the US has partnered with Tennessee Valley Authority's (TVA's) Watts Bar Nuclear Plant and Sequoyah Nuclear Plant to insert tritium-producing burnable absorber rods (TPBARs) that are then processed to extract the tritium~\cite{FR_TVA}.

Furthermore, tritium has a 12.3-year half-life and requires replacement as part of stockpile stewardship. This technical challenge is somewhat addressed in modern weapons that use solid lithium deuteride as fuel in place of some of the tritium: neutrons interact with the lithium-6 and create tritium in situ~\cite{glasstone,krane}. The current treatment of lithium, which can be commonly procured with no known risk to common defense and security, supports an appropriately low level of concern with vertical proliferation triggers.

To the extent proliferation concerns exist with tritium (or lithium) as a nuclear weapons booster, or more particularly, the system that would be responsible for managing or breeding these materials that would be part of a fusion device, this does not mean that a fusion device itself creates a significant proliferation risk requiring application of safeguards or other issues. Instead, the risk is primarily that foreign bad actors abuse this material or underlying breeding technology. Therefore, the materials can be themselves independently secured through export controls. Indeed, both tritium and lithium-6 are subject to strict dual-use export controls administered by the U.S. Department of Commerce, including but not limited to~\cite{EC_DOC,Part_110}:
\begin{itemize}
	\item Export Control Classification Number (ECCN) 1C235, ``Tritium, tritium compounds, mixtures containing tritium in which the ratio of tritium 	to hydrogen atoms exceeds 1 part in 1,000, and 	products or devices containing any of the foregoing,"\footnote{Tritium that is produced specifically in a nuclear reactor is currently controlled by the US NRC in its regulations set forth in 10 CFR Part 110. See ECCN 1C235 note; 10 CFR 110.9(c). The basis for NRC regulation is because the NRC as a general matter regulates the export of any byproduct material produced in a nuclear reactor (including everything from tritium to cesium) regardless if it carries a proliferation concern. See 42 USC 2111, 2112. Indeed, tritium produced through other mechanisms (e.g., particle accelerators) would appear to remain regulated by the Department of Commerce. See ECCN 1C235 note.}
	\item ECCN 1A231, ``Target assemblies and components for the production of tritium," and
	\item ECCN 1C233, ``Lithium enriched in the lithium-6 ($^6$Li) isotope to greater than its natural isotopic abundance, and products or devices containing enriched lithium, as follows: elemental lithium, alloys, compounds, mixtures containing lithium, manufactures thereof, and waste or scrap of any of the foregoing."
\end{itemize} 

These material-based export control protocols are already in place and ready to control the global distribution of tritium and other theoretical boosters when fusion is commercialized. This paper discusses how these controls can be updated to reflect a usage-based ``controls-by-design" approach that provides a high level of assurance that these items cannot be exported to bad actors that will abuse the materials. 

In summary, tritium material does not contribute to the horizontal proliferation of nuclear weapons, but has some potential to contribute to vertical proliferation. That said, vertical proliferation has historically been treated as a lower concern, with multiple sites in the United States containing tritium licensed without any indication of or reference to proliferation risk.  Moreover, the vertical proliferation risk posed by fusion is not particularly novel nor more-enabling than a state already capable of producing, for example, plutonium. To the extent a concern is identified, already existing export controls can control the distribution of these materials.

\section{Conclusion}\label{sec:conclusion}
Understanding that fusion falls outside the NPT and IAEA safeguards framework, as explained in section 2, this paper in section 3 evaluates whether fusion raises a nonproliferation risk substantial enough to warrant amending that framework, or if addressing fusion through dual use export controls is more appropriate. 

Based on the evaluation in this paper, it does not appear that amendment of the NPT is warranted in light of the fundamental differences between the intended use of the fusion technology and its potential contribution to proliferation -- underscored by the fundamental differences between fission and fusion from a nonproliferation standpoint. Additionally, amending the NPT is largely a geopolitical issue and would require extensive global consensus, which is not an easy task and likely infeasible.

With that said, certain components and materials related to fusion are and should continue to be regulated as ``dual-use" items, in alignment with the NSG Part 2 Guidelines,  due to their potential use for commercial and misuse for proliferation purposes. And fusion developers should continue to look to ways to incorporate ``controls by design" into their device design, and engaging in discussions with the nonproliferation community in doing so.  Additional work can be done as the technology matures to develop a right-sized export controls framework for fusion. 
 
Taken together, these measures are and will be sufficient to ensuring only peaceful uses of fusion energy that contribute to global peace, security, and prosperity.

{\scriptsize \bibliography{bibliography}}

\begin{thebibliography}{10}
\newcommand{\enquote}[1]{``#1''}

\bibitem{FIA_survey_2021}
\MakeUppercase{{Fusion Industry Association}} and \MakeUppercase{{UK Atomic
  Energy Authority}}, \enquote{The global fusion industry in 2021,}  (2021).

\bibitem{FIA_survey_2022}
\MakeUppercase{{Fusion Industry Association}}, \enquote{The global fusion
  industry in 2022,}  (2022).

\bibitem{PR_Helion_1}
\MakeUppercase{{Helion Energy}}, \enquote{Helion Energy Achieves 100 Million
  Degrees Celsius Fusion Fuel Temperature and Confirms 16-Month Continuous
  Operation of Its Fusion Generator Prototype,}  (6 2021).

\bibitem{PR_Helion_2}
\MakeUppercase{D.~Kirtley}, \MakeUppercase{A.~Hine}, \MakeUppercase{R.~Milroy},
  \MakeUppercase{C.~Pihl}, \MakeUppercase{R.~Ryan}, \MakeUppercase{A.~Shimazu},
  and \MakeUppercase{G.~Votroubek}, \enquote{Thermonuclear Field Reversed
  Configuration Plasmas in the Trenta Prototype,} \emph{Symposium on Fusion
  Engineering} (12 2021).

\bibitem{PR_Helion_3}
\MakeUppercase{D.~Kirtley}, \MakeUppercase{B.~Campbell},
  \MakeUppercase{A.~Hine}, \MakeUppercase{R.~Milroy}, \MakeUppercase{C.~Pihl},
  and \MakeUppercase{G.~Votroubek}, \enquote{Vacuum Vessel and Divertor Design
  and Results of 16 Month Operation of the Trenta Magneto-Inertial Fusion
  Prototype,} \emph{Symposium on Fusion Engineering} (12 2021).

\bibitem{PR_CFS_1}
\MakeUppercase{{Commonwealth Fusion Systems}}, \enquote{Commonwealth Fusion
  Systems Raises \$1.8 Billion in Funding to Commercialize Fusion Energy,}  (11
  2021).

\bibitem{PR_CFS_2}
\MakeUppercase{{Commonwealth Fusion Systems}}, \enquote{Commonwealth Fusion
  Systems creates viable path to commercial fusion power with world's strongest
  magnet,}  (09 2021).

\bibitem{PR_GF_1}
\MakeUppercase{{General Fusion}}, \enquote{General Fusion Achieves Critical
  Technology Milestone for Practical Fusion Power,}  (01 2022).

\bibitem{PR_TAE_1}
\MakeUppercase{{TAE Technologies}}, \enquote{Claiming a landmark in fusion
  energy, TAE Technologies sees commercialization by 2030,}  (04 2021).

\bibitem{PR_TAE_2}
\MakeUppercase{H.~Gota}, \MakeUppercase{M.~Binderbauer},
  \MakeUppercase{T.~Tajima}, \MakeUppercase{A.~Smirnov},
  \MakeUppercase{S.~Putvinski}, \MakeUppercase{M.~Tuszewski},
  \MakeUppercase{S.~Dettrick}, \MakeUppercase{D.~Gupta},
  \MakeUppercase{S.~Korepanov}, \MakeUppercase{R.~Magee},
  \MakeUppercase{J.~Park}, \MakeUppercase{T.~Roche}, \MakeUppercase{J.~Romero},
  \MakeUppercase{E.~Trask}, \MakeUppercase{X.~Yang},
  \MakeUppercase{P.~Yushmanov}, \MakeUppercase{K.~Zhai},
  \MakeUppercase{T.~DeHaas}, \MakeUppercase{M.~Griswold},
  \MakeUppercase{S.~Gupta}, \MakeUppercase{S.~Abramov},
  \MakeUppercase{A.~Alexander}, \MakeUppercase{I.~Allfrey},
  \MakeUppercase{R.~Andow}, \MakeUppercase{B.~Barnett},
  \MakeUppercase{M.~Beall}, \MakeUppercase{N.~Bolte},
  \MakeUppercase{E.~Bomgardner}, \MakeUppercase{A.~Bondarenko},
  \MakeUppercase{F.~Ceccherini}, \MakeUppercase{L.~Chao},
  \MakeUppercase{R.~Clary}, \MakeUppercase{A.~Cooper}, \MakeUppercase{C.~Deng},
  \MakeUppercase{A.~Dunaevsky}, \MakeUppercase{P.~Feng},
  \MakeUppercase{C.~Finucane}, \MakeUppercase{D.~Fluegge},
  \MakeUppercase{L.~Galeotti}, \MakeUppercase{S.~Galkin},
  \MakeUppercase{K.~Galvin}, \MakeUppercase{E.~Granstedt},
  \MakeUppercase{K.~Hubbard}, \MakeUppercase{I.~Isakov},
  \MakeUppercase{M.~Kaur}, \MakeUppercase{J.~Kinley},
  \MakeUppercase{A.~Korepanov}, \MakeUppercase{S.~Krause},
  \MakeUppercase{C.~Lau}, \MakeUppercase{A.~Lednev},
  \MakeUppercase{H.~Leinweber}, \MakeUppercase{J.~Leuenberger},
  \MakeUppercase{D.~Lieurance}, \MakeUppercase{D.~Madura},
  \MakeUppercase{J.~Margo}, \MakeUppercase{D.~Marshall},
  \MakeUppercase{R.~Marshall}, \MakeUppercase{T.~Matsumoto},
  \MakeUppercase{V.~Matvienko}, \MakeUppercase{M.~Meekins},
  \MakeUppercase{W.~Melian}, \MakeUppercase{R.~Mendoza},
  \MakeUppercase{R.~Michel}, \MakeUppercase{Y.~Mok},
  \MakeUppercase{M.~Morehouse}, \MakeUppercase{R.~Morris},
  \MakeUppercase{L.~Morton}, \MakeUppercase{M.~Nations},
  \MakeUppercase{A.~Necas}, \MakeUppercase{S.~Nicks}, \MakeUppercase{G.~Nwoke},
  \MakeUppercase{M.~Onofri}, \MakeUppercase{A.~Ottaviano},
  \MakeUppercase{R.~Page}, \MakeUppercase{E.~Parke}, \MakeUppercase{K.~Phung},
  \MakeUppercase{G.~Player}, \MakeUppercase{I.~Sato},
  \MakeUppercase{T.~Schindler}, \MakeUppercase{J.~Schroeder},
  \MakeUppercase{D.~Sheftman}, \MakeUppercase{A.~Sibley},
  \MakeUppercase{A.~Siddiq}, \MakeUppercase{M.~Signorelli},
  \MakeUppercase{M.~Slepchenkov}, \MakeUppercase{R.~Smith},
  \MakeUppercase{G.~Snitchler}, \MakeUppercase{V.~Sokolov},
  \MakeUppercase{Y.~Song}, \MakeUppercase{L.~Steinhauer},
  \MakeUppercase{V.~Stylianou}, \MakeUppercase{J.~Sweeney},
  \MakeUppercase{J.~Titus}, \MakeUppercase{A.~Tkachev},
  \MakeUppercase{M.~Tobin}, \MakeUppercase{J.~Ufnal},
  \MakeUppercase{T.~Valentine}, \MakeUppercase{A.~V. Drie},
  \MakeUppercase{J.~Ward}, \MakeUppercase{C.~Weixel}, \MakeUppercase{C.~White},
  \MakeUppercase{M.~Wollenberg}, \MakeUppercase{S.~Ziaei}, \MakeUppercase{the
  TAE~Team}, \MakeUppercase{L.~Schmitz}, \MakeUppercase{Z.~Lin},
  \MakeUppercase{A.~Ivanov}, \MakeUppercase{T.~Asai}, \MakeUppercase{E.~Baltz},
  \MakeUppercase{M.~Dikovsky}, \MakeUppercase{W.~Heavlin},
  \MakeUppercase{S.~Geraedts}, \MakeUppercase{I.~Langmore},
  \MakeUppercase{P.~Norgaard}, \MakeUppercase{R.~V. Behren},
  \MakeUppercase{T.~Madams}, \MakeUppercase{A.~Kast}, and
  \MakeUppercase{J.~Platt}, \enquote{Overview of C-2W: high temperature,
  steady-state beam-driven field-reversed configuration plasmas,} \emph{Nuclear
  Fusion}, \textbf{61}, \emph{10}, 106039 (oct 2021).

\bibitem{PR_TE_1}
\MakeUppercase{{Tokamak Energy}}, \enquote{Tokamak Energy moves closer to
  comemrcial fusion: 100M degree plasma a world record for a spherical
  tokamak,}  (03 2022).

\bibitem{PR_Zap_1}
\MakeUppercase{{Zap Energy}}, \enquote{Zap Energy raises \$27.5 million to
  advance reactor technology,}  (05 2021).

\bibitem{PR_firstlight_1}
\MakeUppercase{{First Light Fusion}}, \enquote{First Light achieves world first
  fusion result, proving unique new target technology,}  (04 2022).

\bibitem{UN_1}
\MakeUppercase{{General Assembly of the United Nations}},
  \enquote{{Establishment of a Commission to Deal with the Problems Raised by
  the Discovery of Atomic Energy},} \emph{Resolutions adopted by the General
  Assembly}, \textbf{1}, \emph{1} (1946).

\bibitem{ICAN}
\MakeUppercase{{International Campaign to Abolish Nuclear Weapons}}.

\bibitem{LANL_primer}
\MakeUppercase{R.~Serber}, \enquote{{The Los Alamos Primer},} \emph{Los Alamos
  National Laboratory Report}, , \emph{LA-1} (1943).

\bibitem{glasstone}
\MakeUppercase{S.~Glasstone} and \MakeUppercase{P.~J. Dolan}, \emph{{The
  Effects of Nuclear Weapons}}, United States Department of Defense and the
  Energy Reseearch and Development Administration, 3 ed. (1977).

\bibitem{krane}
\MakeUppercase{K.~Krane}, \emph{Introductory Nuclear Physics}, Wiley (1991).

\bibitem{history_reactor}
\MakeUppercase{{World Nuclear Association}}.

\bibitem{NPT}
\enquote{Treaty on the Non-Proliferation of Nuclear Weapons,} \emph{United
  Nations Treaty Series}, \textbf{729} (1970).

\bibitem{UNDC}
\enquote{United Nations Disarmament Commission,} \emph{United Nations Office
  for Disarmament Affairs}.

\bibitem{ENDC_1}
\MakeUppercase{A.~Legault}, \MakeUppercase{M.~Fortmann}, and
  \MakeUppercase{D.~Ellington}, \emph{A Diplomacy of Hope: Canada and
  Disarmament, 1945-1988}, McGill-Queen's University Press (1992).

\bibitem{IAEA_glossary_safeguards}
\emph{IAEA Safeguards Glossary: 2001 Edition}, no.~3 in International Nuclear
  Verification Series (2001).

\bibitem{ZanggerCommittee}
\MakeUppercase{{Office for Disarmament Affairs, United Nations}},
  \enquote{Communication dated 18 February 2020 received from the Permanent
  Mission of Denmark regarding the Export of Nuclear Material and of Certain
  Categories of Equipment and Other Material,} \emph{Information Circular of
  the IAEA} (03 2020).

\bibitem{Doyle}
\enquote{Nuclear Safeguards, Security, and Nonproliferation,} in
  \MakeUppercase{J.~E. Doyle}, editor, \enquote{Nuclear Safeguards, Security,
  and Nonproliferation,} Butterworth-Heinemann, Boston, 2nd ed. (2019).

\bibitem{PANDA}
\MakeUppercase{D.~Reilly}, \MakeUppercase{N.~Ensslin}, and
  \MakeUppercase{H.~Smith}, \emph{Passive Nondestructive Assay of Nuclear
  Materials}, National Technical Information Service (1991).

\bibitem{pandendum}
\MakeUppercase{N.~Ensslin}, \MakeUppercase{W.~H. Geist}, \MakeUppercase{M.~S.
  Krick}, and \MakeUppercase{M.~M. Pickrell}, \emph{Passive Nondestructive
  Assay of Nuclear Materials Addendum, Chapter 7: Active Neutron Multiplicity
  Counting} (2007).

\bibitem{ensslin_guide}
\MakeUppercase{D.~G. Langner}, \MakeUppercase{J.~E. Stewart},
  \MakeUppercase{M.~M. Pickrell}, \MakeUppercase{M.~S. Krick},
  \MakeUppercase{N.~Ensslin}, and \MakeUppercase{W.~C. Harker},
  \enquote{Application Guide to Neutron Multiplicity Counting,} .

\bibitem{IAEA_Legal}
\MakeUppercase{L.~Rockwood}, \enquote{Legal frameworks for IAEA safeguards,}
  (2013).

\bibitem{IAEA_CSA}
\MakeUppercase{{International Atomic Energy Agency}}, \enquote{{The Structure
  and Content of Agreements Between the Agency and States Required in
  Connection with the Treaty on the Non-Proliferation of Nuclear Weapons},}
  \emph{IAEA Information Circulars}, \textbf{153} (1972).

\bibitem{IAEA_China}
\MakeUppercase{{International Atomic Energy Agency}}, \enquote{{Protocol
  Additional to the Agreement Between the People's Republic of China and the
  International Atomic Energy Agency for the Application of Safeguards in
  China},} \emph{IAEA Information Circulars}, \textbf{369} (2002).

\bibitem{IAEA_France}
\MakeUppercase{{International Atomic Energy Agency}}, \enquote{{Protocol
  Additional to the Agreement between France, the European Atomic Energy
  Community and the International Atomic Energy Agency for the Application of
  Safeguards in France},} \emph{IAEA Information Circulars}, \textbf{290}
  (2005).

\bibitem{IAEA_Russia}
\MakeUppercase{{International Atomic Energy Agency}}, \enquote{{Protocol
  between the Russian Federation and the International Atomic Energy Agency
  Additional to the Agreement between the Union of Soviet Socialist Republics
  and the International Atomic Energy Agency for the Application of Safeguards
  in the Union of Soviet Socialist Republics},} \emph{IAEA Information
  Circulars}, \textbf{327} (2008).

\bibitem{IAEA_UK}
\MakeUppercase{{International Atomic Energy Agency}}.

\bibitem{IAEA_US}
\MakeUppercase{{International Atomic Energy Agency}}, \enquote{{Protol
  Additional to the Agreement between the United States of America and the
  International Atomic Energy Agency for the Application of Safeguards in the
  United States of America},} \emph{IAEA Information Circulars}, \textbf{288}
  (2009).

\bibitem{IAEA_ItemSpecific}
\MakeUppercase{{International Atomic Energy Agency}}.

\bibitem{SQP}
\MakeUppercase{{International Atomic Energy Agency}}, \enquote{Safeguards
  Implementation Guide for States with Small Quantities Protocols,} \emph{IAEA
  Services Series}, \textbf{22} (2013).

\bibitem{AdditionalProtocol}
\enquote{Model Protocol Additional to the Agreement(s) Between State(s) and the
  International Atomic Energy Agency for the Application of Safeguards,}
  \emph{International Atomic Energy Agency}, \textbf{INFCIRC/540 (Corrected)}
  (1997).

\bibitem{AdditionalProtocol_NRC}
\MakeUppercase{{Nuclear Regulatory Commission}}, \enquote{Additional Protocol
  to the U.S. -- IAEA Safeguards Agreement,}  (Accessed 27 April 2022).

\bibitem{PubLaw_Nonpro}
\enquote{{Nuclear Non-Proliferation Act of 1978},} \emph{Public Law},
  \textbf{95-242}, \emph{92}, 120 (03 1978).

\bibitem{PubLaw_AEA}
\enquote{{Atomic Energy Act of 1954},} \emph{Public Law}, \textbf{83-703}
  (1954).

\bibitem{EC_NRC1}
\enquote{{Component and other parts of facilities},} \emph{U.S. Code Title 42},
  \textbf{Chapter 23}, \emph{Division A, Subchapter IX, \S 2139}.

\bibitem{EC_NRC2}
\enquote{{Export and Import of Nuclear Equipment and Material},} \emph{Federal
  Register}, \textbf{43}, \emph{98} (05 1978).

\bibitem{Part_110}
\MakeUppercase{{Nuclear Regulatory Commission}}, \enquote{{10 CFR Part 110:
  Export and Import of Nuclear Equipment and Material},}  (Page Last Updated:
  December 30, 2021).

\bibitem{EC_NRC3}
\enquote{{Assistance to Foreign Atomic Energy Activities},} \emph{Federal
  Register}, \textbf{80}, \emph{35} (02 2015).

\bibitem{Part_810}
\enquote{{10 CFR Part 810: Assistance to Foreign Atomic Energy Activities},}
  \emph{Code of Federal Regulations} (Page Last Updated: May 20, 2022).

\bibitem{Part_730}
\enquote{{15 CFR Part 730: General Information},} \emph{Code of Federal
  Regulations} (Page Last Updated: May 13, 2022).

\bibitem{PubLaw_EC_1979}
\enquote{{Export Administration Act of 1979},} \emph{Public Law},
  \textbf{96-72}, \emph{93} (09 1979).

\bibitem{PubLaw_EC_2018}
\enquote{{Export Control Reform Act of 2018},} \emph{Public Law},
  \textbf{115-232}, \emph{132} (08 2018).

\bibitem{EC_DOC}
\MakeUppercase{{Department of Commerce}}, \enquote{Category 1 - Special
  Materials and Related Equipment, Chemicals, ``Microorganisms," and
  ``Toxins",} \emph{Commerce Control List} (2021).

\bibitem{FR_Deuterium}
\enquote{{Updates on the Export of Deuterium},} \emph{Federal Register},
  \textbf{86}, \emph{191}, 55476--55479 (10 2021).

\bibitem{Part_738}
\enquote{{15 CFR Part 730: Commerce Control List Overview and the Country
  Chart},} \emph{Code of Federal Regulations} (Page Last Updated: May 13,
  2022).

\bibitem{NPT_noFusion1}
\MakeUppercase{{Office of Public Communication, Bureau of Public Affairs,
  United States}}, \enquote{The Department of State Bulletin,} \textbf{58}
  (1968).

\bibitem{NPT_noFusion2}
\MakeUppercase{{United States Arms Control and Disarmament Agency}},
  \enquote{Documents on Disarmament,}  (09 1969).

\bibitem{NPT_Germany}
\MakeUppercase{{Office for Disarmament Affairs, United Nations}},
  \enquote{Treaty on the Non-Proliferation of Nuclear Weapons,}
  \emph{Depository Notes} ({Last Updated 2010}).

\bibitem{NPT_Germany2}
\MakeUppercase{{Office for Disarmament Affairs, United Nations}},
  \enquote{Germany: Ratification of Treaty on the Non-Proliferation of Nuclear
  Weapons (NPT),} \emph{Depository Notes} (1975).

\bibitem{NPT_Book}
\MakeUppercase{M.~Shaker}, \enquote{The Nuclear Non-Proliferation Treaty:
  Origins and Implementation, 1959-1979,} \emph{Oceana Publications, Inc.}
  (1980).

\bibitem{Part_74}
\MakeUppercase{{Nuclear Regulatory Commission}}, \enquote{{10 CFR Part 74:
  Material Control and Accounting of Special Nuclear Material},}  (Page Last
  Updated: September 23, 2020).

\bibitem{Part_73}
\MakeUppercase{{Nuclear Regulatory Commission}}, \enquote{{10 CFR Part 73:
  Physical Protection of Plants and Materials},}  (Page Last Updated: December
  30, 2021).

\bibitem{fusion_breeder}
\MakeUppercase{R.~W. Moir}, \enquote{The fusion breeder,} \emph{Journal of
  Fusion Energy}, \textbf{2}, 351--367 (1982).

\bibitem{goldston}
\MakeUppercase{R.~J. Goldston} and \MakeUppercase{A.~Glaser},
  \enquote{{Safeguard Requirements for Fusion Power Plants},} \emph{Princeton
  Plasma Physics Laboratory Report}, \textbf{4794} (8 2012).

\bibitem{glaser}
\MakeUppercase{A.~Glaser} and \MakeUppercase{R.~Goldston},
  \enquote{Proliferation risks of magnetic fusion energy: clandestine
  production, covert production and breakout,} \emph{Nuclear Fusion},
  \textbf{52}, \emph{4}, 043004 (mar 2012).

\bibitem{fissionFusionHybrid}
\MakeUppercase{B.~Kuteev}, \MakeUppercase{E.~Azizov}, \MakeUppercase{A.~Bykov},
  \MakeUppercase{A.~Dnestrovsky}, \MakeUppercase{V.~Dokuka},
  \MakeUppercase{G.~Gladush}, \MakeUppercase{A.~Golikov},
  \MakeUppercase{P.~Goncharov}, \MakeUppercase{M.~Gryaznevich},
  \MakeUppercase{M.~Gurevich}, \MakeUppercase{A.~Ivanov},
  \MakeUppercase{R.~Khairutdinov}, \MakeUppercase{V.~Khripunov},
  \MakeUppercase{D.~Kingham}, \MakeUppercase{A.~Klishchenko},
  \MakeUppercase{V.~Kurnaev}, \MakeUppercase{V.~Lukash},
  \MakeUppercase{S.~Medvedev}, \MakeUppercase{P.~Savrukhin},
  \MakeUppercase{V.~Sergeev}, \MakeUppercase{Y.~Shpansky},
  \MakeUppercase{A.~Sykes}, \MakeUppercase{G.~Voss}, and
  \MakeUppercase{A.~Zhirkin}, \enquote{Steady-state operation in compact
  tokamaks with copper coils,} \emph{Nuclear Fusion}, \textbf{51}, \emph{7},
  073013 (may 2011).

\bibitem{fissionFusionHybrid2}
\MakeUppercase{P.~Bagryansky}, \MakeUppercase{A.~Ivanov},
  \MakeUppercase{E.~Kruglyakov}, \MakeUppercase{A.~Kudryavtsev},
  \MakeUppercase{Y.~Tsidulko}, \MakeUppercase{A.~Andriyash},
  \MakeUppercase{A.~Lukin}, and \MakeUppercase{Y.~Zouev}, \enquote{Gas dynamic
  trap as high power 14 MeV neutron source,} \emph{Fusion Engineering and
  Design}, \textbf{70}, \emph{1}, 13--33 (2004), international Science and
  Technology Center: Fusion Projects.

\bibitem{PPPL_1995}
\MakeUppercase{V.~Finley} and \MakeUppercase{M.~Wieczorek}, \enquote{Princeton
  Plasma Physics Laboratory Annual Site Environmental Report for Calendar Year
  1995,} \textbf{PPPL-3284} (1998).

\bibitem{PPPL_1996}
\MakeUppercase{V.~Finley} and \MakeUppercase{J.~Levine}, \enquote{Princeton
  Plasma Physics Laboratory Annual Site Environmental Report for Calendar Year
  1996,} \textbf{PPPL-3290} (1998).

\bibitem{PPPL_1998}
\MakeUppercase{V.~Finley}, \enquote{Princeton Plasma Physics Laboratory Annual
  Site Environmental Report for Calendar Year 1998,} \textbf{PPPL-3440} (2000).

\bibitem{PPPL_1999}
\MakeUppercase{V.~Finley}, \enquote{Princeton Plasma Physics Laboratory Annual
  Site Environmental Report for Calendar Year 1999,} \textbf{PPPL-3557} (2001).

\bibitem{PPPL_2000}
\MakeUppercase{V.~Finley}, \enquote{Princeton Plasma Physics Laboratory Annual
  Site Environmental Report for Calendar Year 2000,} \textbf{PPPL-3688} (2002).

\bibitem{PPPL_2001}
\MakeUppercase{V.~Finley}, \enquote{Princeton Plasma Physics Laboratory Annual
  Site Environmental Report for Calendar Year 2001,} \textbf{PPPL-3938} (2004).

\bibitem{PPPL_2002}
\MakeUppercase{V.~Finley}, \enquote{Princeton Plasma Physics Laboratory Annual
  Site Environmental Report for Calendar Years 2002 and 2003,}
  \textbf{PPPL-4039} (2004).

\bibitem{PPPL_2004}
\MakeUppercase{V.~Finley}, \enquote{Annual Site Environmental Report for
  Calendar Year 2004,} \textbf{PPPL-4396} (2009).

\bibitem{PPPL_2005}
\MakeUppercase{V.~Finley}, \enquote{Annual Site Environmental Report for
  Calendar Years 2005-2006,} \textbf{PPPL-4481} (2010).

\bibitem{Part_141}
\MakeUppercase{{Environmental Protection Agency}}, \enquote{{40 CFR Part 141:
  National Primary Drinking Water Regulations},}  (2015).

\bibitem{NRC_FIA}
\MakeUppercase{{Fusion Industry Association}}, \enquote{Offiste Impacts of
  Fusion: Normal Operation / Off-Normal Shutoff,} \emph{NRC Public Meeting:
  Developing a Regulatory Framework for Commercial Fusion Energy Systems}
  (2022).

\bibitem{Part_742}
\enquote{{15 CFR Part 742: Control Policy -- CCL Based Controls},} \emph{Code
  of Federal Regulations} (Page Last Updated: May 13, 2022).

\bibitem{CNN_tractors}
\MakeUppercase{O.~Fylyppov} and \MakeUppercase{T.~Lister}, \enquote{Russians
  plunder \$5M farm vehicles from Ukraine -- to find they've been remotely
  disabled,} \emph{Cable News Network (CNN)} (2022).

\bibitem{Craig}
\MakeUppercase{S.~O. O'Neal} and \MakeUppercase{J.~Clark}, \enquote{Microsoft
  and Open AI Comment on Advance Notice of Proposed Rulemaking (ANPRM) for the
  Identification and Review of Controls for Certain Foundational Technologies,}
   (2020).

\bibitem{NAS_ICF}
\MakeUppercase{N.~R. Council}, \emph{Assessment of Inertial Confinement Fusion
  Targets}.

\bibitem{NIF}
\MakeUppercase{J.~A. Paisner}, \MakeUppercase{E.~M. Campbell}, and
  \MakeUppercase{W.~J. Hogan}, \enquote{The National Ignition Facility
  Project,} \emph{Fusion Technology}, \textbf{26}, \emph{3P2}, 755--766 (1994).

\bibitem{ZMachine_1}
\MakeUppercase{M.~K. Matzen} and \MakeUppercase{et~al.},
  \enquote{Pulsed-power-driven high energy density physics and inertial
  confinement fusion research,} \emph{Physics of Plasmas}, \textbf{12},
  \emph{5}, 055503 (2005).

\bibitem{ZMachine_2}
\MakeUppercase{M.~Savage}, \MakeUppercase{K.~LeChien},
  \MakeUppercase{W.~Stygar}, \MakeUppercase{J.~Maenchen},
  \MakeUppercase{D.~McDaniel}, and \MakeUppercase{K.~Struve}, \enquote{Overview
  and Status of the Upgraded Z Pulsed Power Driver,} in \enquote{2008 IEEE
  International Power Modulators and High-Voltage Conference,}  (2008), pp.
  93--93.

\bibitem{ZMachine_3}
\MakeUppercase{D.~B. Sinars} and \MakeUppercase{et~al.}, \enquote{Review of
  pulsed power-driven high energy density physics research on Z at Sandia,}
  \emph{Physics of Plasmas}, \textbf{27}, \emph{7}, 070501 (2020).

\bibitem{ICF_NPT_Lasers}
\MakeUppercase{O.~o. A.~C. US~DOE} and \MakeUppercase{Nonproliferation},
  \enquote{The National Ignition Facility (NIF) and the Issue of
  Nonproliferation: Final Study,}  (1995).

\bibitem{ICF_CTBT_Clinton}
\MakeUppercase{S.~Jone}, \MakeUppercase{R.~Kidder}, and \MakeUppercase{F.~von
  Hippel}, \enquote{The Question of Pure-Fusion Explosions under the CTBT,}
  \textbf{51}, \emph{9} (1998).

\bibitem{ICF_declassification}
\MakeUppercase{O.~o.~D. US~DOE}, \enquote{Restricted Data Declassification
  Decisions, 1946 to the Present,}  (2001).

\bibitem{MCNP6}
\MakeUppercase{C.~J. Werner}, \MakeUppercase{J.~S. Bull}, \MakeUppercase{C.~J.
  Solomon}, \MakeUppercase{F.~B. Brown}, \MakeUppercase{G.~W. McKinney},
  \MakeUppercase{M.~E. Rising}, \MakeUppercase{D.~A. Dixon},
  \MakeUppercase{R.~L. Martz}, \MakeUppercase{H.~G. Hughes},
  \MakeUppercase{L.~J. Cox}, \MakeUppercase{A.~J. Zukaitis},
  \MakeUppercase{J.~C. Armstrong}, \MakeUppercase{R.~A. Forster}, and
  \MakeUppercase{L.~Casswell}, \enquote{{MCNP} Version 6.2 Release Notes,} .

\bibitem{DOE_tritium}
\MakeUppercase{{US DOE}}, \enquote{{Restricted Data Declassification Decisions,
  1946 to the Present (RDD-8)},}  (1 2002).

\bibitem{FR_TVA}
\enquote{{Production of Tritium for the United States Department of Energy,
  Rhea and Hamilton Counties, TN},} \emph{Federal Register}, \textbf{65},
  \emph{88}, 26259 (5 2000).

\bibitem{FR_37_2}
\enquote{{Physical Protection of Category 1 and Category 2 Quantities of
  Radioactive Material},} \emph{Federal Register}, \textbf{81}, \emph{49} (3
  2016).

\bibitem{Part_37}
\MakeUppercase{{Nuclear Regulatory Commission}}, \enquote{{10 CFR Part 37:
  Physical Protection of Category 1 and Category 2 Quantities of Radioactive
  Material},}  (Page Last Updated: September 15, 2021).

\bibitem{IAEA_Code_of_Conduct}
\MakeUppercase{IAEA}, \enquote{{Code of Conduct on the Safety and Security of
  Radioactive Sources},}  (2004).

\bibitem{FR_37}
\enquote{{Appendix A to Part 37 -- Category 1 and Category 2 Radioactive
  Materials},} \emph{Federal Register}, \textbf{78}, \emph{53}, 17020 (3 2013).

\bibitem{G8}
\enquote{{Non Proliferation of Weapons of Mass Destruction, Securing
  Radioactive Sources},} \emph{G8 Summit} (2003).

\bibitem{IAEA_security}
\MakeUppercase{IAEA}, \enquote{{Security of Radioactive Sources},}
  \emph{Proceedings of the International Conference on Security of Radioactive
  Sources} (2003).

\bibitem{TECDOC_1344}
\MakeUppercase{IAEA}, \enquote{{Categorization of radioactive sources},}
  \emph{IAEA-TECDOC-1344} (2003).

\end{thebibliography}
\bibliographystyle{ans}

\appendix
\section{Use of tritium in radiological dispersal devices}\label{sec:RDD}
A potential concern raised with fusion is the use of certain associated radioactive materials in radiological dispersal devices (RDDs). These RDDs or, ``dirty bombs," are different from nuclear weapons: instead of utilizing nuclear reactions to achieve large detonation power, a conventional explosive is used to spread radioactive material across a populated area.  All radioactive materials, such as cobalt-60, are potentially usable in an RDD. In the fusion context, tritium (T) -- a fusion fuel or byproduct for most designs -- could theoretically be included in an RDD in the form of tritiated water (HTO), directly, or in the elemental form of HT if the explosive device causes the HT to oxidize to HTO\footnote{The dose consequences/health effects of HTO are $\approx$10,000 greater than that of HT. Additionally, HT rises much like a helium-filled balloon, whereas HTO does not and has a higher probability of entering the ecosystem.}. Note that RDDs are not considered a proliferation concern (rather a broader security concern), and is included here for completeness.

Note that fusion facilities produce neutrons that may activate (make radioactive) materials via neutron interactions, and these materials may reach quantities and concentrations relevant to RDD risk. Quantity and concentration depend on the fusion power plant design (materials to be activated and their proximity to the fusion), the power level (amount of neutrons), and fuel type (amount of neutrons and neutron energy -- higher-energy neutrons, such as those from deuterium-tritium fusion (14.1 MeV) compared to those from deuterium-deuterium fusion (2.5 MeV), have more ways to activate material). Since (i) the treatment of activated materials would be equivalent to that of tritium, and (ii) the range of materials is specific to power plants, specific analysis is omitted in this paper. 

The use of radioactive materials in RDDs is not a novel concern, nor does it raise nuclear weapon proliferation issues given that the materials used cannot themselves be used as the core of a nuclear weapon. Defense against RDDs is a matter of (physical) security rather than safeguards. The current countermeasures (for over 80,000 sources~\cite{FR_37_2}) in the US are set forth by the Nuclear Regulatory Commission (NRC) via its regulations in 10 CFR Part 37: Physical Protection of Category 1 and Category 2 Quantities of Radioactive Material~\cite{Part_37}. Part 37 provides ``requirements for the physical protection program for any licensee that possesses an aggregated category 1 or category 2 quantity of radioactive material," with specific radioactive materials of concern delineated in an appendix to the regulations. Category 1 and 2 materials are classified as follows, where $A$ is activity and $D$ is an isotope-specific danger metric:
\begin{itemize}
	\item Category 1 sources, if not safely managed or securely protected, would likely cause permanent injury to a person in contact with them for more than a few minutes.  It would likely be fatal to be close to this amount of unshielded material for a period of a few minutes to an hour. The activity ratio (A/D) is $\geq$ 1000. These sources are typically used in radiothermal generators, irradiators, and radiation teletherapy~\cite{Part_37,IAEA_Code_of_Conduct}.
	
	\item Category 2 sources, if not safely managed or securely protected, could cause permanent injury to a person in contact with them for a short time (minutes to hours). It could be fatal to be close to this amount of unshielded radioactive material for a period of hours to days. The activity ratio is 1000 $>$ (A/D) $\geq$ 10. These sources are typically used in industrial gamma radiography, high- and medium-dose rate brachytherapy, and radiography~\cite{Part_37,IAEA_Code_of_Conduct}.
\end{itemize}

The thresholds for category 1 and category 2 are based on the IAEA Code of Conduct titled ``Code of Conduct on the Safety and Security of Radioactive Sources"~\cite{FR_37,IAEA_Code_of_Conduct}. This code reflects the findings produced by the International Conference on Security of Radioactive Sources held in Vienna in March 2003 (the Hofburg Conference). The G-8 annual summit (of ``the Heads of State and Government of the eight major industrialised democracies and the Representatives of the European Union") then issued a statement on ``non proliferation of weapons of mass destruction -- securing radioactive sources" wherein it encouraged all countries to observe the Code of Conduct by strengthening controls on radioactive sources~\cite{G8}. 

The NRC has determined that its regulations, based off of the IAEA Code of Conduct, are sufficient to provide reasonable protection against RDDs by ``protecting these materials from theft or diversion"~\cite{Part_37}. Note that plutonium \textit{is} included in the category listing. Part 37 contains a robust set of security protections, including access controls, monitoring and detection, establishment of security zones, and coordination with local law enforcement. Additional protections exist for the transport of radioactive materials. 

Currently, tritium is not listed in the NRC's listing of category 1 and category 2 materials, but inserting tritium would not be difficult. Tritium was left out of the listing because it was considered ``never available in sufficient quantities at one location to be of use in an RDD" ~\cite{IAEA_security}.  Thus, if the advent of fusion energy sufficiently changes this assumption on the abundance of tritium, it is appropriate and sufficient (insofar as RDDs) to maintain the security of tritium via Part 37. 

Tritium-specific threshold values for category 1 and category 2 would be based on D-values that define a dangerous source, i.e., ``a source that could, if not under control, give rise to exposure sufficient to cause severe deterministic effects." IAEA TECDOC-1344 derives the D-values~\cite{TECDOC_1344} and also ``provides the underlying methodology that could be applied to radionuclides not listed"~\cite{IAEA_Code_of_Conduct}. Tritium is listed in Table I.2 of TECDOC-1344. Based on these values (and assuming a tritium specific activity of 9640 Ci/g), the category 1 threshold for tritium would be $2\times10^6$ TBq ($\approx5\times10^7$ Ci, $\approx$5.6 kg) and the threshold for category 2 would be $2\times10^4$ TBq ($\approx5\times10^5$ Ci, $\approx$56 g). Note that isotope thresholds are smaller when combined with other radioactive materials and the levels are determined by a sum of fractions law (e.g., if a site possesses 25\% of the cobalt-60 threshold and no other radioisotopes, the tritium threshold would be 75\% of the aforementioned values).

Taken together, while the presence of tritium in significant quantities poses a security risk by way of RDDs, it is not a nuclear weapon proliferation risk. The US regulatory regime via 10 CFR Part 37 addresses existing RDD risks through physical security and monitoring measures in place for category 1 and category 2 quantities of radioactive material. The tritium RDD risk can be handled by augmenting Part 37, particularly Table 1 in Appendix A, with the appropriate quantity thresholds for tritium: 56 g for category 2 and 5.6 kg for category 1. 

\end{document}